\documentclass[preprints,article,accept,moreauthors,pdftex]{Definitions/mdpi} 

\firstpage{1} 
\pubvolume{xx}
\issuenum{1}
\articlenumber{5}
\pubyear{2019}
\copyrightyear{2019}
\history{Received: date; Accepted: date; Published: date}

\Title{Fully Depleted Monolithic Active Microstrip Sensors: TCAD simulation study of an innovative design concept}

\Author{Lorenzo De Cilladi $^{1,2,}$*\orcidA{}, Thomas Corradino $^{3,4}$, Gian-Franco Dalla Betta $^{3,4}$, Coralie Neub{\"u}ser $^{4}$ and Lucio Pancheri $^{3,4}$ on behalf of the ARCADIA collaboration}

\AuthorNames{Firstname Lastname, Firstname Lastname and Firstname Lastname}

\address{%
$^{1}$ \quad Dipartimento di Fisica, Università degli Studi di Torino, 10125 Torino (Italy)\\
$^{2}$ \quad Istituto Nazionale di Fisica Nucleare (INFN), Sezione di Torino, 10125 Torino (Italy)\\
$^{3}$ \quad Dipartimento di Ingegneria Industriale, Università degli Studi di Trento, 38123 Trento (Italy)\\
$^{4}$ \quad Trento Institute for Fundamental Physics and Applications - Istituto Nazionale di Fisica Nucleare (TIFPA-INFN), 38123 Trento (Italy)}

\corres{Correspondence: lorenzo.decilladi@unito.it}

\abstract{
The paper presents the simulation studies of 10\,$\mu$m pitch microstrips on a fully depleted monolithic active CMOS technology and describes their potential to provide a new and cost-effective solution for particle tracking and timing applications. The Fully Depleted Monolithic Active Microstrip Sensors (FD-MAMS) described in this work, which are developed within the framework of the ARCADIA project, are compliant with commercial CMOS fabrication processes. A TCAD simulation campaign was performed in the perspective of an upcoming engineering production run with the aim of designing FD-MAMS, studying their electrical characteristics and optimising the sensor layout for enhanced performance in terms of low capacitance, fast charge collection and low-power operation. A very fine pitch of 10\,$\mu$m was chosen to provide very high spatial resolution. This small pitch still allows readout electronics to be monolithically integrated in the inter-strip regions, enabling the segmentation of long strips and the implementation of distributed readout architectures. The effects of surface radiation damage expected for total ionising doses of the order of 10 to 10$^5$\,krad were also modelled in the simulations. The results of the simulations exhibit promising performance in terms of timing and low power consumption and motivate R\&D efforts to further develop FD-MAMS; the results will be experimentally verified through measurements on the test structures that will be available at the beginning of 2021.}

\keyword{Particle detectors; silicon detectors; monolithic sensors; microstrip sensors; CMOS; TCAD simulations; fast timing.}

\begin{document}


\section{Introduction}
\label{section:introduction}

Charged particle tracking and timing are fundamental tools for both physics research and for numerous applications. Although a number of detection techniques are available, silicon detectors have become largely employed due to their versatility and to the parallel strong developments of the semiconductor industry. Various flavours of silicon sensors have been developed to meet the specific requirements of different experiments and applications, such as high spatial resolution, fast charge collection, low power consumption, high radiation tolerance and low cost per unit area.

Silicon detectors are divided in two categories, namely hybrid and monolithic detectors. The former are made of two separate silicon elements, the sensor and the chip, which are interconnected through external bump or wire bonding. While the sensor hosts the sensing volume only, the chip integrates the front-end readout electronics. On the contrary, monolithic sensors, which are emerging as a valid alternative to hybrid detectors, embed the front-end electronics in the same silicon substrate which hosts the sensing volume, with benefits in terms of material budget, production yield and fabrication cost, as they are produced with commercial microelectronics processes~\cite{wermes2009pixel, wermes2019pixel, garcia2018review}.

Due to their characteristics, monolithic sensors have recently raised a wide interest in different research fields; studies, proposals and developments have been made for applications in high energy physics (HEP)~\cite{mager2016alpide, pernegger2017first, wang2017development}, X-ray imaging~\cite{wunderer2014percival, hatsui2013direct}, medical particle imaging~\cite{mattiazzo2018impact} and space experiments~\cite{scotti2019high}.\\

The state of the art includes three main types of monolithic sensors. The first type, called Depleted Field Effect Transistors (DEPFETs), is capable of low noise operation thanks to low sensor input capacitance~\cite{rummel2009intrinsic}. DEPFET detectors have been developed and used for HEP applications~\cite{marinas2011belle}, for X-ray imaging in space~\cite{treis2010mixs} and for free electron laser experiments~\cite{lutz2010depfet}. The main limitation of DEPFETs is the need to reset their internal gate which can be quickly saturated by the leakage~\cite{lutz2016depfet}, thus making this technology not suitable for environments with high levels of non-ionising radiation.

A second approach consists in the SOI (Silicon-On-Insulator) monolithic sensors. SOI sensors embed a buried-oxide layer separating a thin low-resistivity silicon layer, which hosts the integrated readout circuitry, from a thicker high-resistivity substrate, which serves as the sensitive detection region~\cite{kucewicz2005development, lan2015soi}. This technology allows a low capacitance to be obtained~\cite{lan2015soi}; however, SOI sensors suffer from back-gate effect and have a reduced radiation hardness, due to accumulation of positive holes charges in the buried oxide layer after irradiation~\cite{hagino2020radiation}. Strategies have been found to overcome these limitations and to recover from the Total Ionising Dose (TID)~\cite{asano2016characteristics}, but, as a consequence, the fabrication process of SOI sensors have become highly specialised and not compliant with standard microelectronics production processes. This results in increased cost per unit area, which is a critical issue for large-area detector applications.

A third flavour of monolithic sensors is represented by CMOS sensors~\cite{turchetta2003monolithic}. CMOS sensors were already in use for light detection when they were first proposed for charged particle tracking at the beginning of the 2000s~\cite{turchetta2001monolithic}. Over the last years, important advancements in CMOS sensors allowed them to be employed in many applications, eventually leading to very large scale productions for particle trackers at collider experiments. The STAR pixel detector, which took data at the Relativistic Heavy Ion Collider (RHIC) from 2014 to 2016, was the first large area monolithic pixel tracker ever built, for a total of 0.16\,m$^2$~\cite{contin2018star}. These dimensions have been exceeded by the newly-constructed Inner Tracking System of the ALICE experiment at CERN, in which a total detector surface of about 10\,m$^2$ is covered by ALPIDE CMOS monolithic active\footnote{A monolithic sensors is called “active” if it integrates a signal amplifier inside each pixel or strip.} pixel sensors (MAPS)~\cite{mager2016alpide}.

These achievements demonstrate the level of maturity and reliability that CMOS sensors have recently reached. However, there is still room for further improvements, especially in terms of charge collection speed and radiation hardness, and possibility to push previous limits in terms of low power density, high spatial resolution and SNR.\\

Pixel detectors are the first choice for small scale applications and for vertex trackers at collider experiments~\cite{wermes2009pixel} as they have an intrinsic capability of providing a two-dimensional position information~\cite{rossi2006pixel}. On the other hand, microstrip sensors~\cite{peisert1992silicon} are largely used as particle detectors for space applications and are a competitive option for particle trackers due to their high spatial resolution, simpler readout and much lower power density (i.e. power consumption per unit area) compared to pixel detectors.
Large experiments at particle colliders have largely employed silicon hybrid strip sensors in the past, and are still developing and assembling new large-area trackers based on this technology, 
as in the case of the Phase-2 Upgrades of the CMS Outer Tracker~\cite{chowdhury2020phase} and of the ATLAS Strip Inner Tracker~\cite{david2018new}. Recent space experiments equipped with silicon hybrid microstrip trackers include FERMI-LAT~\cite{atwood2007design}, DAMPE~\cite{azzarello2016dampe}, PAMELA~\cite{adriani2003magnetic} and AMS-02~\cite{lubelsmeyer2011upgrade}. Strip-like sensors integrated in a monolithic technology have been proposed by combining the outputs of 55\,$\mu$m\,$\times$\,55$\mu$m~\cite{benhammadi2019decal} or 40\,$\mu$m\,$\times$\,600$\mu$m~\cite{han2020study} pixels in each column or row of a pixel matrix.

Spatial resolution of 1.25-1.3\,$\mu$m was achieved using hybrid silicon microstrip sensors with 25\,$\mu$m pitch~\cite{straver1994one}. However, it has recently been demonstrated with fully depleted double-SOI monolithic pixel sensors that the 1\,$\mu$m limit can be exceeded by semiconductor detectors~\cite{sekigawa2017fine}. The keys to a high spatial resolution with analogue readout are a fine microstrip pitch, a low sensor thickness to reduce Coulomb scattering and delta-ray emission, and an increased SNR, which can be achieved by reducing the leakage current and the sensor input capacitance to the readout electronics, but which is ultimately limited by the noise of the front-end electronics~\cite{turchetta1993spatial, peisert1992silicon, straver1994one}.\\

This paper presents the first investigation, design and simulation studies of CMOS Fully Depleted Monolithic Active Microstrip Sensors with 10\,$\mu$m pitch for charged particle detection. Properly optimised sensor layouts may allow sub-micron resolution, improved radiation hardness and fast timing performance thanks to full depletion~\cite{wang2017development, snoeys2017process} in a power-saving and cost-effective commercial technology. Moreover, a further advantage of monolithic microstrips is the potential complexity reduction of the detector assembly compared to hybrid microstrip detectors. In fact, since many readout functions can be monolithically integrated on the same chip which hosts the sensing volume, 1-by-1 strip bonding to the external readout electronics would not be needed anymore. We have hence studied and designed the FD-MAMS within the framework of the INFN ARCADIA project in order to provide an innovative solution for satellite-based space trackers and for large area particle detectors at future collider experiments.

The results of the Technology Computer-Aided Design (TCAD) simulation campaign\footnote{The TCAD simulations were produced using the Synopsys \textsuperscript{\textregistered} Sentaurus (Version O-2018.06-SP2) software.} which allowed different MAMS design flavours to be compared in terms of sensor capacitance, reference voltage values, leakage current and charge collection time and efficiency are presented; the effects of the inclusion of a silicon dioxide (SiO$_2$) layer on top of the sensor and of surface radiation damage on the sensor operating parameters are explored; the study of charge sharing between groups of adjacent strips when particles with different Linear Energy Transfer (LET) traverse the sensor is reported. A selection of the MAMS presented in this paper is going to be implemented in test structures which were submitted in November 2020 for an engineering production run.\\

The paper is organised as follows: Section~\ref{section:materials_methods} presents the sensor concept for the ARCADIA fully depleted CMOS monolithic microstrip sensors and illustrates the simulation campaign that was performed for the sensor design optimisation; Section~\ref{section:results} describes and discusses the results of the simulations; Section~\ref{section:conclusions} presents the conclusions, the future perspectives and the planned tests for the ARCADIA monolithic microstrip sensors.


\section{The ARCADIA sensor concept}
\label{section:materials_methods}

The ARCADIA project and its precursor, SEED (Sensor with Embedded Electronics Development), designed an innovative sensor concept~\cite{pancheri2019110, pancheri2020fully} based on a modified 110\,nm CMOS process developed in collaboration with LFoundry and compatible with their standard 110\,nm CMOS process. Up to 6 metal layers can be stacked on top of the sensor, for a total metal and insulator thickness of about 4-5\,$\mu$m. The ARCADIA collaboration is developing a scalable event-driven readout architecture to cover large detection surfaces (O(cm$^2$)) while maintaining ultra-low power consumption. The target for pixel sensors is 10-20\,mW/cm$^2$ at high rates O(100\,MHz), but for less dense particle environments (e.g. in space applications) a dedicated low-power operation mode implements a cyclic pulling of the data packets from each section of the pixel matrix and disables most of the serialisers and data transceivers, further reducing the total power consumption of the chip.

\begin{figure}[H]
\centering
\includegraphics[width=11 cm]{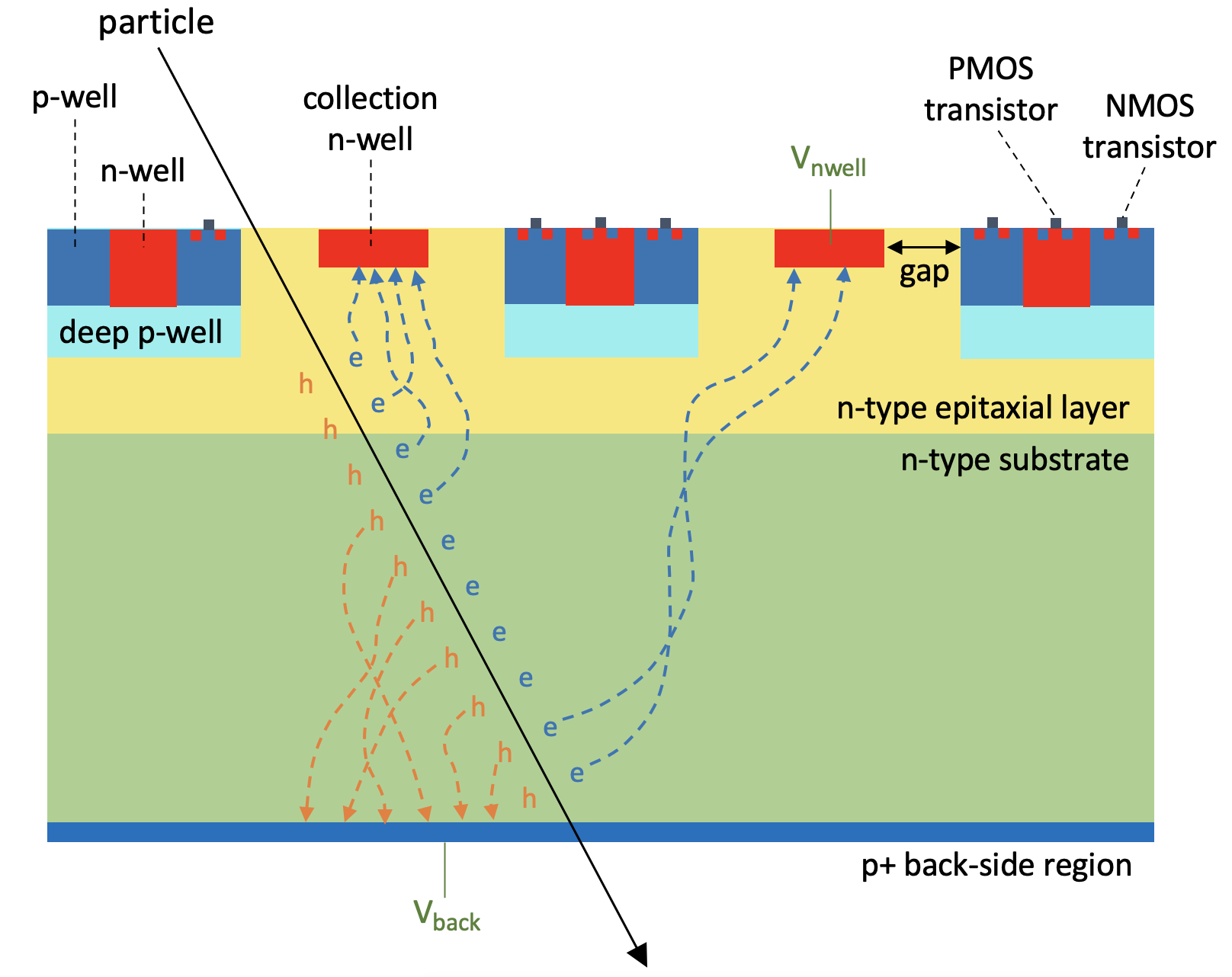}
\caption{ARCADIA monolithic sensor concept. The dotted arrows indicate the drift path of electrons (e) and holes (h) generated by a particle crossing the sensor. The voltages V$_{nwell}$ and V$_{back}$ applied to the sensor contacts are shown in green.}
\label{fig:sensor_concept}
\end{figure}

In our project, an n-on-n sensor concept enabling full substrate depletion over tens or hundreds of microns and allowing full CMOS electronics to be implemented was employed. A simplified view of the sensor cross section is visible in Figure~\ref{fig:sensor_concept}. The process allows to achieve sensor thicknesses from 50 to 400\,$\mu$m. A high resistivity n-type substrate was used and constitutes the active volume. The sensing n-well node, located on top of the sensor, collects the electrons produced by ionisation due to particles traversing the active detection volume.

N-doped and p-doped wells intended to host pMOSFETs and nMOSFETs respectively are shielded by a deep p-well, which allows the integration of full CMOS electronics and, hence, more complex digital functions, when necessary. In fact, the deep p-well prevents the n-wells hosting pMOSFETs from competing with the n-doped sensing node in the collection of the charge, thus avoiding loss of charge collection efficiency.

\vspace{-0.6cm}

A p+ boron-doped region sits at the backside of the n-substrate, thus forming a pn-junction; when a negative bias voltage V$_{back}$ is applied to the backside p+ contact, sensor depletion starts from the pn-junction at the bottom of the sensor and eventually extends to the whole sensor, if the backside voltage is sufficiently large. Since the high voltage needed for sensor depletion is applied at the backside, it is possible to maintain the voltage V$_{nwell}$ applied to the front n-well electrode below 1\,V and to use low-voltage integrated electronics (1.2\,V transistors) which is more radiation-resistant and has lower noise. Full sensor depletion allows fast charge collection by drift (beneficial to enhance the timing performance), higher charge collection efficiency, deeper collection depth and larger SNR; it also leads to improved radiation tolerance, as charge losses by trapping are reduced~\cite{pernegger2017first}. Since thicker sensors need higher backside bias voltage to reach full depletion, termination structures composed of multiple floating guard rings are used to avoid early breakdown at the edges of the pn-junction.

An additional n-type epitaxial layer, with lower resistivity than the substrate, is integrated between the n-type substrate and the deep p-wells. Its aim is to better control the potential barrier below the deep p-well, in order to delay the onset of the punch through current described in details in Paragraph~\ref{subsubsec:punch_through}.\\

The feasibility of this sensor concept and approach to Fully Depleted monolithic CMOS sensors was proven in the framework of the SEED project~\cite{pancheri2019110, pancheri2020fully}. The upcoming ARCADIA engineering run will include different design flavours of FD-CMOS monolithic sensors, both pixelated and strip-like. Large-area ($1.3 \times 1.3$\,cm$^2$) pixel demonstrators with embedded CMOS electronics and pixel test structures ($0.5 \times 0.5$ and $1.5 \times 1.5$\,mm$^2$) without integrated readout circuitry~\cite{neubueser2020sensor} are foreseen, with pitches ranging from 10 to 50\,$\mu$m. The test structures will include as well the innovative MAMS and will allow a detailed characterisation of these sensors. The 3D TCAD simulation campaign performed to design the first FD-MAMS will be presented and discussed in the following.

\subsection{TCAD simulations}
\label{subsec:tcad_simulations}

3D TCAD simulations were employed as a tool to optimise the sensor layout and performance. The use of 3D simulations is necessary to have a more realistic domain and results which are more accurate and less affected by boundary conditions. Furthermore, we were also interested in studying the charge collection dynamics after a particle crosses the sensor, and this is more straightforward with 3D simulations. A fine pitch of 10\,$\mu$m was chosen for the microstrips in order to explore the characteristics and performance of a sensor layout which pushes the requirements on both spatial and timing resolution. Different sensor thicknesses foreseen for the production runs were simulated. Variations in the sensor layout and operating parameters were tested to study and optimise the sensor response.
The simulated sensor flavours take into account the limitations imposed by the foundry’s sensors fabrication process, especially for the n-well and p-well sizes. The strip simulations investigated sensor flavours which pushed the design to the limits of the process requirements.

\vspace{-0.2cm}
\begin{figure}[H]
\centering
\includegraphics[width=\textwidth]{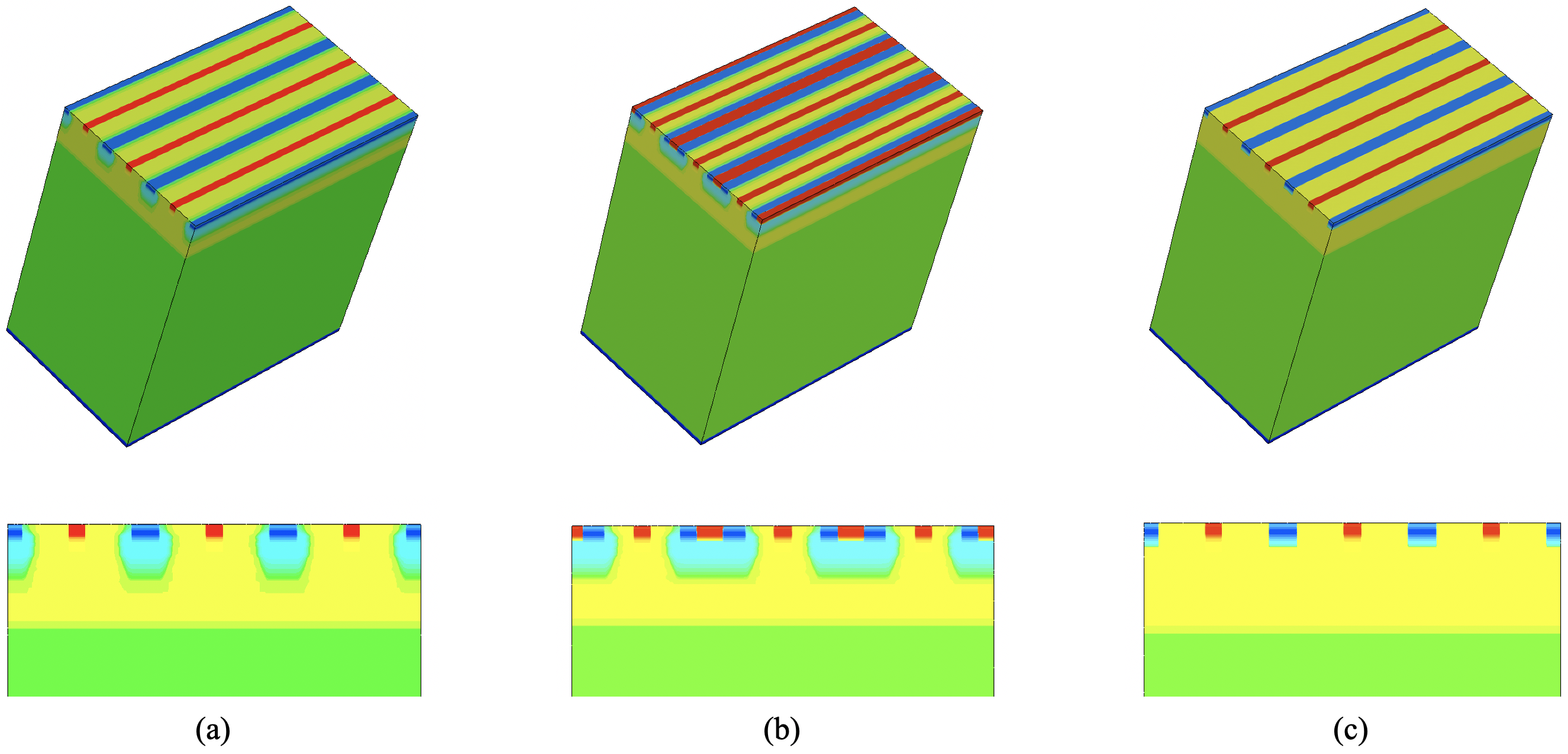}
\caption{Example TCAD 3D sensor domains for ARCADIA microstrips (top row) and corresponding cross sections (bottom row). (a) Standard simulation domain for sensors with the deep p-well. (b) Addition of n-wells above the deep p-wells. (c) Simulated ARCADIA microstrips without deep p-wells.}
\label{fig:3strip_domain}
\end{figure}

All the TCAD simulations were performed at a temperature of 300\,K. A standard simulation domain including three 50\,$\mu$m long, 50\,$\mu$m thick, 10\,$\mu$m pitch microstrips is shown as an example in Figure~\ref{fig:3strip_domain} (a). The n-doped substrate is shown in green, the epitaxial layer in yellow, the microstrip sensing n-wells in red, the p-wells in blue and the less doped deep p-well in light blue. The default value for V$_{nwell}$ is 0.8\,V. The p-wells, instead, are kept at a voltage V$_{pwell} = 0$\,V.

One of the simulated sensor flavours has been specifically designed to allow for CMOS digital library cells to be integrated along the strips and is shown in Figure~\ref{fig:3strip_domain} (b). This sensor variant would allow the deployment of complex CMOS digital functions along the strip for distributed signal processing. We observed that the n-wells dedicated to the implementation of PMOS transistors and shielded by the deep p-well do not significantly influence the electrical characteristics of the detector in the TCAD simulation results. Therefore, we did not include them in the simulations.

The deep p-well can be removed in the test structures that will be used to characterise the sensor (see Figure~\ref{fig:3strip_domain}, c), and the necessary CMOS front-end electronics can be deployed at the end of the strips in the chip periphery. Sensors without the deep p-well were simulated as well.

Different n-well, p-well and deep p-well sizes were considered to find the optimal layout in terms of sensor performance. Simulations were also employed to predict the effects that possible production uncertainties can have on the sensor operating parameters and electrical characteristics. For instance, the thickness and resistivity of the epitaxial layer may vary within a confidence range around their typical specified value (see Appendix~\ref{appendix:expected_epi_effect}). 3D simulations for the different cases were run and compared.
Some simulation parameters were fine-tuned using characterization results from a previous set of test structures, produced in the framework of the SEED project~\cite{pancheri2019110}.

\subsection{Electrical and transient simulation}
\label{subsec:electrical_transient_sim}

In this Section, the simulations performed to extract the sensor electrical characteristics and to study the charge collection dynamics are briefly illustrated. Shared definitions and conventions on simulation setups and operating parameters were agreed for the whole ARCADIA simulation campaign and are also described in~\cite{neubueser2020sensor}. The strip length in the upcoming production run will be 1.2\,cm. However, MAMS with lengths of 50\,$\mu$m were simulated in order to run a large set of TCAD simulations in a reasonable computational time. The results were then scaled to the desired length.

\subsubsection{Depletion voltage}
\label{subsubsec:depletion}

\begin{figure}[H]
\centering
\includegraphics[width=9.5 cm]{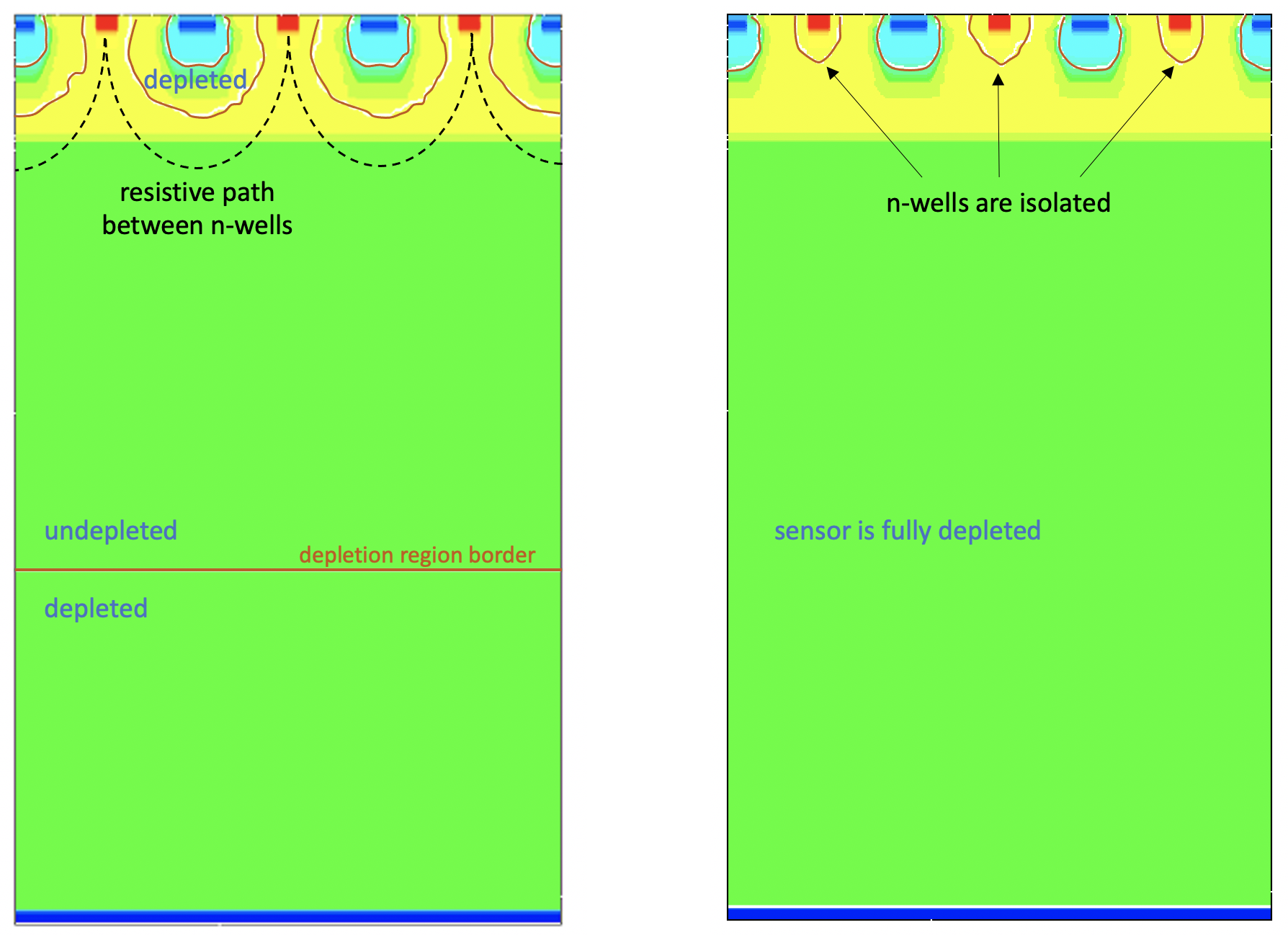}
\caption{Depletion process in ARCADIA microstrips. On the left: cross section of a sensor before full depletion is reached. On the right: cross section of a fully depleted MAMS. The orange lines indicate the edge of the depletion region.}
\label{fig:depletion_process}
\end{figure}

Sensor depletion starts at the backside, where the pn-junction between the n-type substrate and the p+ contact is located. If no negative bias voltage is applied to the backside contact, the sensor is not fully depleted and the collection n-wells are not isolated. This means that a resistive path exists between the n-type sensing nodes (see Figure~\ref{fig:depletion_process}, on the left). Therefore, if a voltage difference is applied between two adjacent n-wells, a current will flow between them.

As the negative voltage applied to the backside contact increases, the space charge region enlarges through the high resistivity substrate, eventually merging with the depletion volume which surrounds the pn-junctions formed between the n-type substrate or epitaxial layer and the deep p-wells. At this point, the sensor is fully depleted, the resistive path between the sensing nodes is closed and the collection n-wells are isolated; this is shown in Figure~\ref{fig:depletion_process}, on the right. In this condition, no current (except for the leakage current) will flow among adjacent n-wells even when different voltages are applied to them.

\vspace{0.2cm}

\begin{figure}[H]
\centering
\includegraphics[width=9.5 cm]{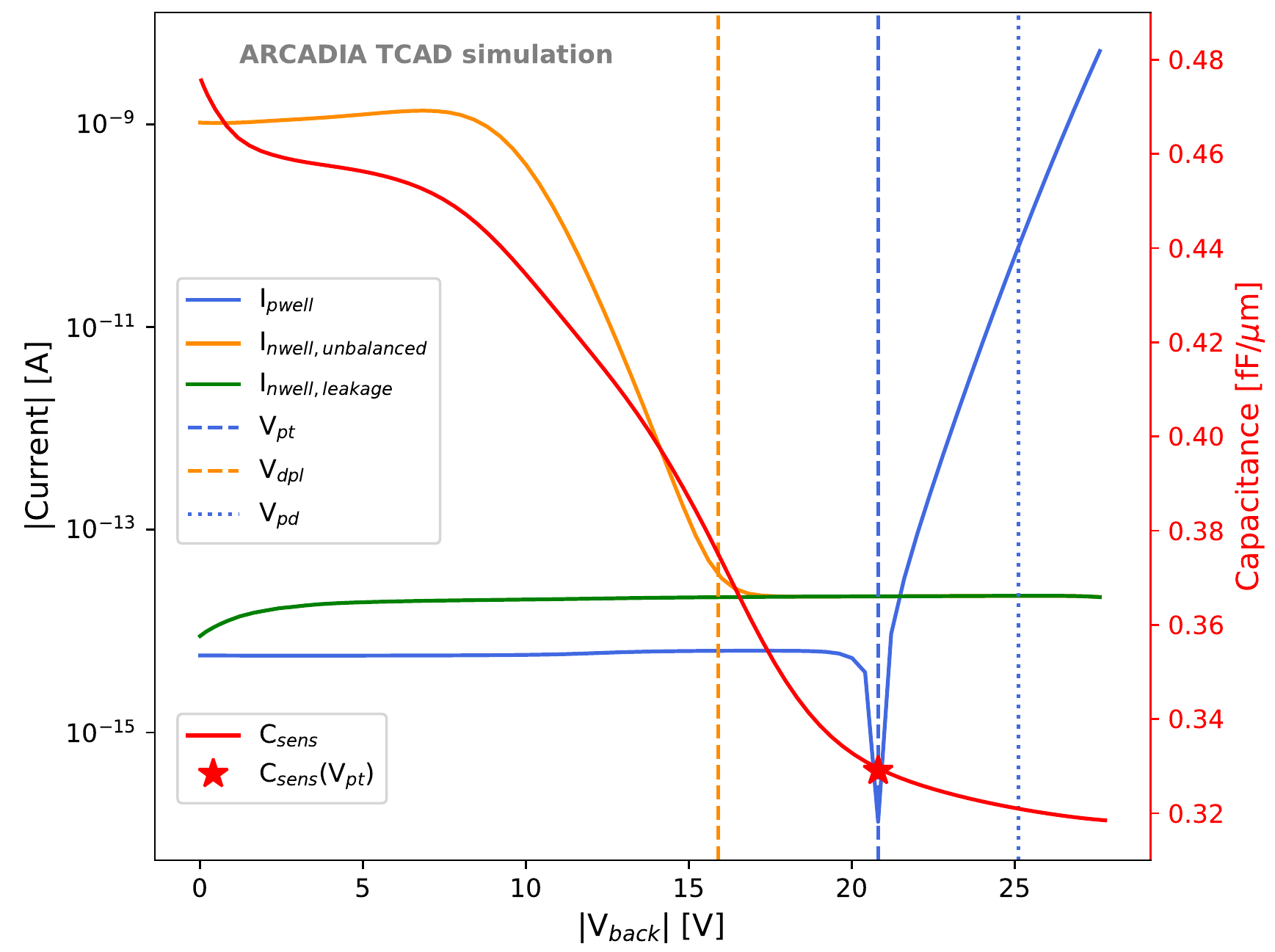}
\caption{Example characteristic IV and CV curves extracted from TCAD simulations of ARCADIA monolithic sensors. The red vertical axis refer to the sensor capacitance ($C_{sens}$) CV curve.}
\label{fig:iv_cv}
\end{figure}

This behaviour can be observed in the orange example IV curve in Figure~\ref{fig:iv_cv} ($I_{nwell, unbalanced}$). The simulated domain shown in Figure~\ref{fig:3strip_domain} (left) was used. In this simulation, a voltage unbalance of 10\,mV was applied between adjacent strips: the first n-well was biased at 0.79\,V, the central one at 0.8\,V and the third one at 0.81\,V. The curve shows the current measured at the sensing node of the central strip as a function of $|V_{back}|$. A current of about 1\,nA is measured at $V_{back} = 0$\,V. As the backside voltages increases and the space charge region enlarges, the current starts decreasing, and eventually reaches a plateau at a current of about $10^{-5}$\,nA. This baseline corresponds to the leakage current (green IV curve, $I_{nwell, leakage}$). The backside voltage at which the single microstrips become isolated and the plateau is reached is the sensor depletion voltage $V_{dpl}$; this voltage is evaluated as the intersection point between the exponential decay fitting of the IV curve decreasing segment and the baseline.

\vspace{0.1cm}

\begin{figure}[H]
\centering
\includegraphics[width=13.4 cm]{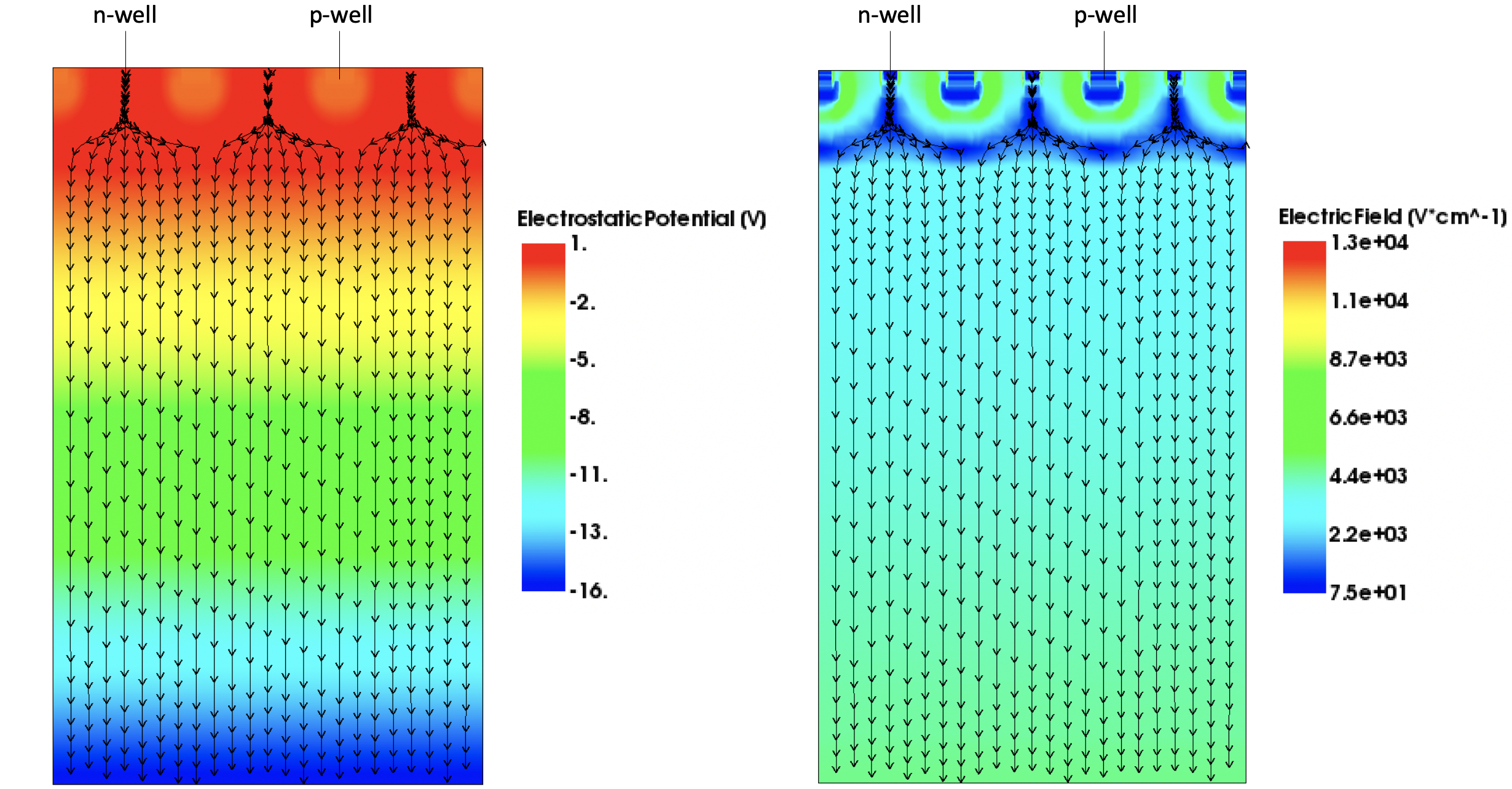}
\caption{Electrostatic potential map (left) and electric field map (right) for a group of three ARCADIA microstrip sensors at $V_{back} = V_{dpl}$. The electric field lines are plotted on top of both maps.}
\label{fig:potential_Efield_lines}
\end{figure}

Figure~\ref{fig:potential_Efield_lines} shows the simulated electrostatic potential and electric field maps at $V_{back} = V_{dpl}$ in a cross section of a 3-strip domain with all the n-wells at $V_{nwell} = 0.8$\,V. Electric field lines are plotted on top of both the electrostatic potential and the electric field maps.

\vspace{-0.3cm}

\subsubsection{Punch-through}
\label{subsubsec:punch_through}

If $V_{back}$ exceeds a certain value, a hole current flowing between the shallow p-doped backside region and the (deep) p-well exponentially increases. This condition is known as punch-through and the hole current is the punch-through current~\cite{chu1972thermionic}. We define the voltage corresponding to the onset of the punch-through as $V_{pt}$. The onset of the punch-through currents can be observed from the blue IV curve in Figure~\ref{fig:iv_cv} ($I_{pwell}$), which shows the absolute value of the current measured at the top p-well contacts as a function of $|V_{back}|$. The dip in the curve, corresponding to the point of sign inversion of the current, was defined as $V_{pt}$. The simulation domain includes three 50\,$\mu$m long, 50\,$\mu$m thick, 10\,$\mu$m pitch microstrips. In this case, the n-wells are all biased at $V_{nwell} = 0.8$\,V, which is the default value.

Sensor operation in low punch-through regime can be tolerated, whereas a too large punch-through current ought to be avoided, as it determines a substantial increase in the power consumption of the whole detector. For this reason, we chose $V_{back}$\,=\,$V_{pt}$ as a safe reference sensor operating voltage; this is the operating point for all the results shown in the following, if not stated differently. The sensor power density can be defined as $pd = \frac{V_{back} \cdot (I_{pwell} + I_{nwell})}{A}$, where $I_{nwell}$ and $I_{pwell}$ are the currents flowing at the sensing node and at the top p-well contacts respectively, and $A$ is the top surface area of the simulated microstrip domain. In order to quantify the maximum acceptable backside bias voltage that limits the absorbed power density, the value $V_{pd}$ at which $pd = 0.1$\,mW/cm$^2$ was extracted from the simulated IV curves (see Figure~\ref{fig:iv_cv}).

\begin{figure}[H]
\centering
\includegraphics[width=12.6 cm]{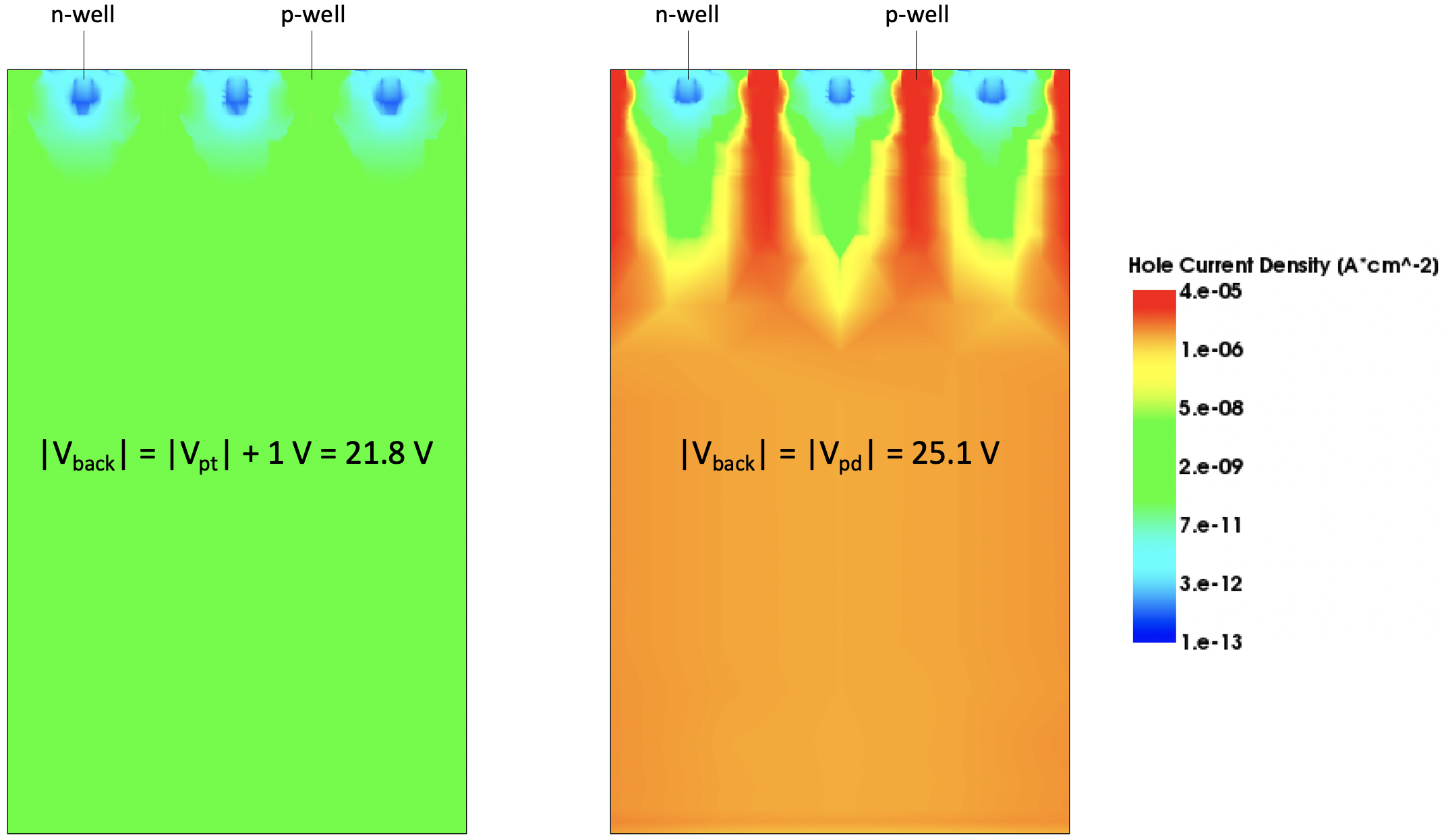}
\caption{Hole current density in a simulated sensor domain including three microstrips in punch-through condition at two different $V_{back}$.}
\label{fig:punch_through}
\end{figure}

Figure~\ref{fig:punch_through} shows the hole current density at two different $|V_{back}|$\,>\,$|V_{pt}|$ in the simulation domain used to extract the $I_{pwell}$ curve of Figure~\ref{fig:iv_cv}. On the left, a backside voltage exceeding $V_{pt}$ by 1\,V was chosen, while on the right $V_{back}$ was set to $V_{pd}$. An increase in the hole current density of several orders of magnitude can be observed below the deep p-wells and in the substrate.

Care had to be taken to ensure that $|V_{dpl}|$\,<\,$|V_{pt}|$ in the designed sensors. In this way, full depletion is reached before the onset of the punch-through. Moreover, the voltage operating range between $V_{dpl}$ and $V_{pt}$, defined as $\Delta V_{op} = |V_{pt} - V_{dpl}|$, should be large enough to ensure safe operation in full depletion before the onset of the punch-through even if deviations from the simulated design occur in the sensor fabrication process.

\subsubsection{Leakage current}
\label{subsubsec:leakage_current}

The same sensor domain and n-well voltage configuration used for the extraction of $V_{pt}$ was also used to evaluate the sensor leakage current $I_{leak}$. The leakage current is defined as the current flowing at the collection nodes in full depletion and in absence of external stimuli, such as particles or radiation. The leakage current as a function of the backside bias voltage is shown in Figure~\ref{fig:iv_cv} as a green curve ($I_{leak}$). In the example shown in Figure~\ref{fig:iv_cv}, a value of 10\,fA was extracted for $I_{leak}$ at $V_{back} = V_{pt}$.

\subsubsection{Sensor capacitance}
\label{subsubsec:sensor_capacitance}

The sensor CV curve was simulated through AC simulations with a frequency of 10\,kHz using the same sensor domain employed for $V_{pt}$ and $I_{leak}$ evaluation, with $V_{nwell} = 0.8$\,V. The major contribution to the sensor capacitance $C_{sens}$, which is the input capacitance seen by the DC-coupled front-end electronics, originates from the lateral capacitance between the collection n-well and the surrounding p-wells. It is thus important to minimize this contribution by a careful selection of the distance between the edge of the collection n-well and the p-wells; we call this distance "gap" (see Figure~\ref{fig:sensor_concept}). An example CV curve is shown in red in Figure~\ref{fig:iv_cv}, with the capacitance per unit length considered. In the example of Figure~\ref{fig:iv_cv}, a value of about 0.33\,fF/$\mu$m was obtained at $V_{back} = V_{pt}$.

It has to be mentioned that in these sensors the depletion voltage does not necessarily correspond to the voltage of minimum capacitance. The reason for this is the presence of the epitaxial layer, which is located far from the backside pn-junction and has a lower resistivity than the substrate. Therefore, the depletion of the epitaxial layer begins after the depletion of the substrate and progresses more slowly with voltage. Full depletion of the whole sensor, including the epitaxial layer, and minimum capacitance are only reached at $|V_{back}| > |V_{dpl}|$. From this point, both capacitance and leakage current values will be intended at $V_{back} = V_{pt}$.

A central focus of the layout optimisation was the minimisation of the sensor capacitance. In fact, low input capacitance to the DC-coupled CMOS readout electronics allows for low-noise readout, low analog power~\cite{pernegger2017first} and, in particular, SNR maximisation. Large input capacitance worsens the noise levels and the speed of the front-end electronics~\cite{wang2017development}.

\subsubsection{Surface radiation damage}
\label{subsubsec:sio2_radiation}

In the simulation campaign performed to study the properties of MAMS, a silicon dioxide (SiO$_2$) layer was added on the top-side of the sensor. In addition to this, surface damage was modeled to evaluate the effects of Total Ionising Dose (TID) on the sensor electrical properties.

\begin{figure}
\centering
\includegraphics[width=9.5 cm]{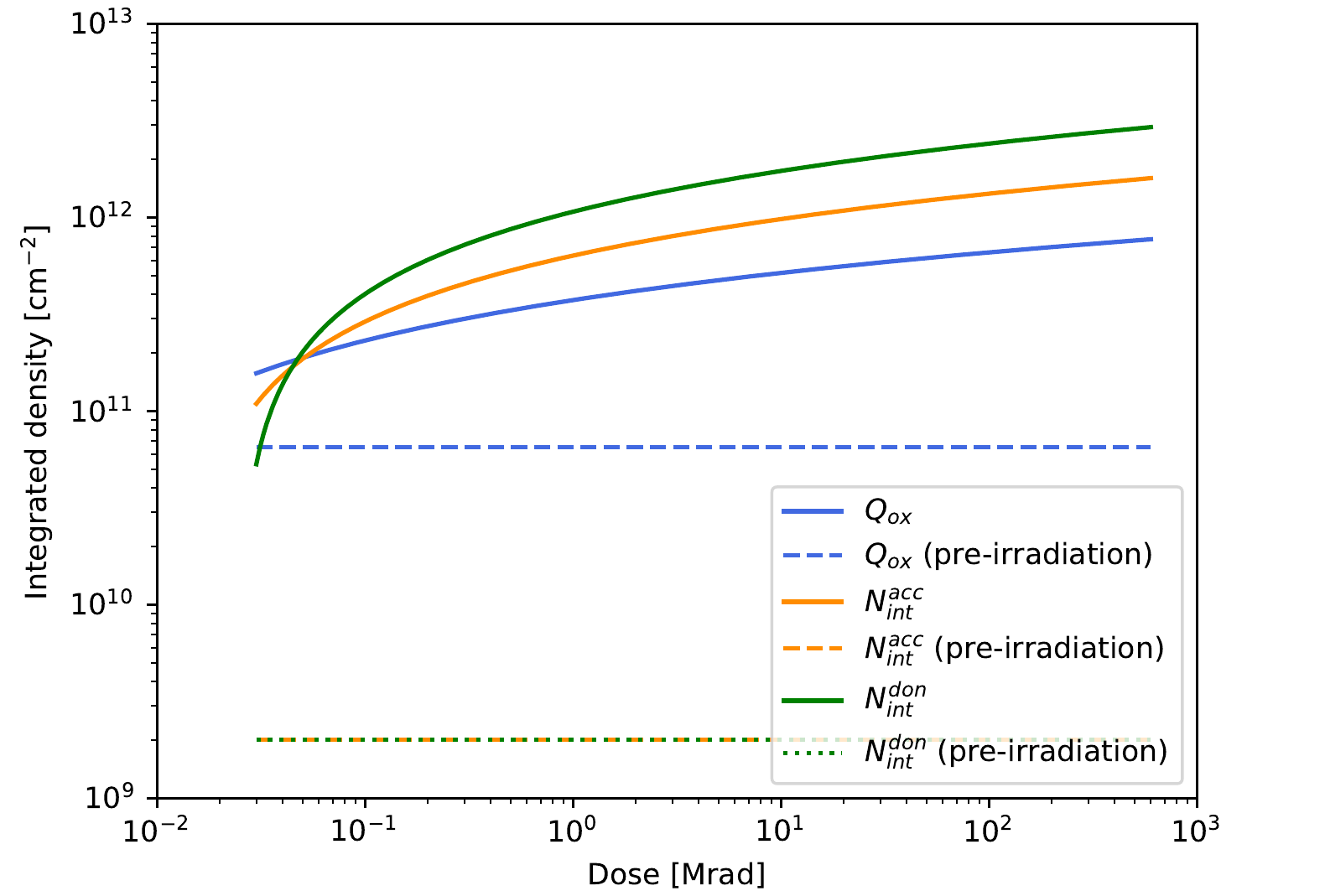}
\caption{Dependence of the oxide charge density $Q_{ox}$, acceptor integrated interface trap state density $N_{int}^{acc}$ and donor integrated interface trap state density $N_{int}^{don}$ on the dose for the surface radiation damage model described in~\cite{passeri2019tcad}. Pre-irradiation values are shown as horizontal dotted lines.}
\label{fig:aida_model}
\end{figure}

The impact of surface radiation damage was modeled following the AIDA-2020-D7.4 report~\cite{passeri2019tcad}. The model introduces fixed positive oxide charges and band-gap acceptor/donor defect levels (trap states) at the Si-SiO$_2$ interface. The concentrations of oxide charges and defect levels start from a fixed value before irradiation (i.e. with the only inclusion of the SiO$_2$ surface layer, at $dose = 0$) and increase with the dose provided to the sensors. The dependence of the oxide charge density $Q_{ox}$~[charges\,$\cdot$\,cm$^{-2}$], of the acceptor integrated interface trap state density $N_{int}^{acc}$~[cm$^{-2}$] and of the donor integrated interface trap state density $N_{int}^{don}$~[cm$^{-2}$] on the dose is shown in Figure~\ref{fig:aida_model}. Pre-irradiation values, shown as dotted horizontal lines in Figure~\ref{fig:aida_model}, are $Q_{ox} = 6.5 \cdot 10^{10}$\,charges\,$\cdot$\,cm$^{-2}$, $N_{int}^{acc} = 2.0 \cdot 10^{9}$\,cm$^{-2}$ and $N_{int}^{don} = 2.0 \cdot 10^{9}$\,cm$^{-2}$. 

In the simulation campaign, the effects of the inclusion of the SiO$_2$ layer and of the radiation damage on the leakage current, sensor capacitance, depletion voltage and punch-through voltage were investigated and will be discussed in Section~\ref{section:results}.

\subsubsection{Transient simulations}
\label{subsubsec:transient_sim}

TCAD transient simulations were run to study the sensor charge collection process in response to particles traversing the simulated microstrip domain. These simulations also let us identify the most relevant layout parameters to be optimised for improving the sensor performance in terms of fast and uniform charge collection irrespective of the particle incidence position. The transient simulations employ the Synopsys \textsuperscript{\textregistered} Sentaurus TCAD \textit{HeavyIon} model, described in~\cite{sentaurus2018sdevice}. The \textit{HeavyIon} model gives an analytical description of the amount of charge generated within a 3D cylindrical distribution along the incident particle track. Two main parameters have to be passed to the \textit{HeavyIon} model: the linear Energy Transfer (LET), defined as the average deposited charge per unit length, and the transverse size of the charge deposition volume generated around the particle trajectory. We chose the charge transverse distribution profile to be gaussian around the particle track.

\begin{figure}[H]
\centering
\includegraphics[width=12.7 cm]{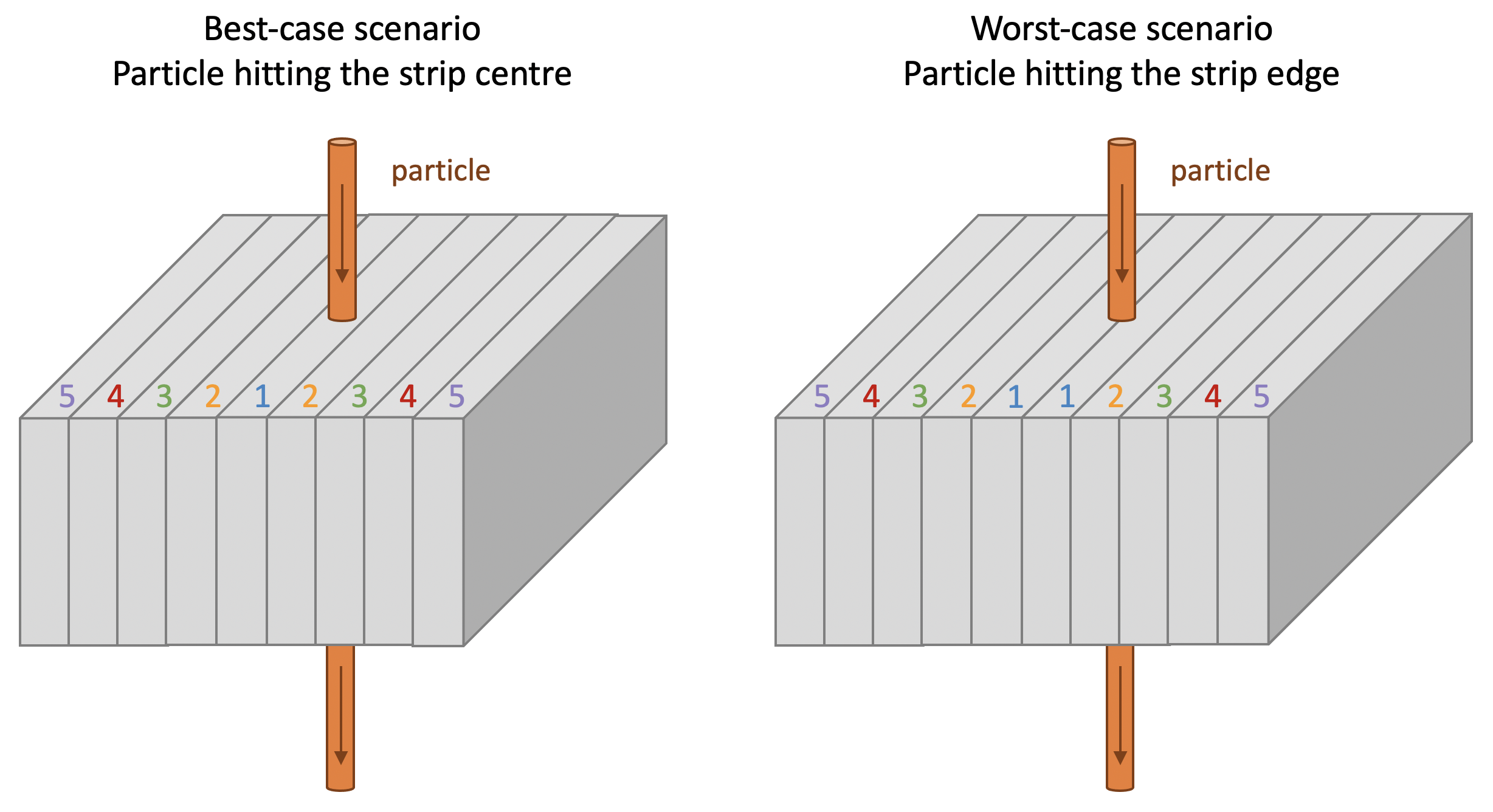}
\caption{Best-case and worst-case scenarios considered in the TCAD transient simulations. The microstrips are represented as adjacent grey blocks and the particle traversing the domain is shown as an orange cylinder. The nomenclature used to identify the microstrips (from 1 to 5) is illustrated.}
\label{fig:best_worst}
\end{figure}

Two extreme cases in terms of particle impact position were studied to evaluate the uniformity of charge collection time and charge collection efficiency. Particle trajectories perpendicular to the sensor surface were considered. In the best-case scenario, the particle impact point corresponds to the centre of a microstrip, which is the centre of a collection n-well. On the contrary, in the worst-case scenario, the particle traverses the sensor at the edge between two adjacent microstrips, i.e. in the middle of a p-well. In Figure~\ref{fig:best_worst} the two cases and the corresponding numbering of the strips are illustrated. This conventional strip nomenclature will be used in the following when referring to transient simulations.\\

\begin{figure}[H]
\centering
\includegraphics[width=5.2 cm]{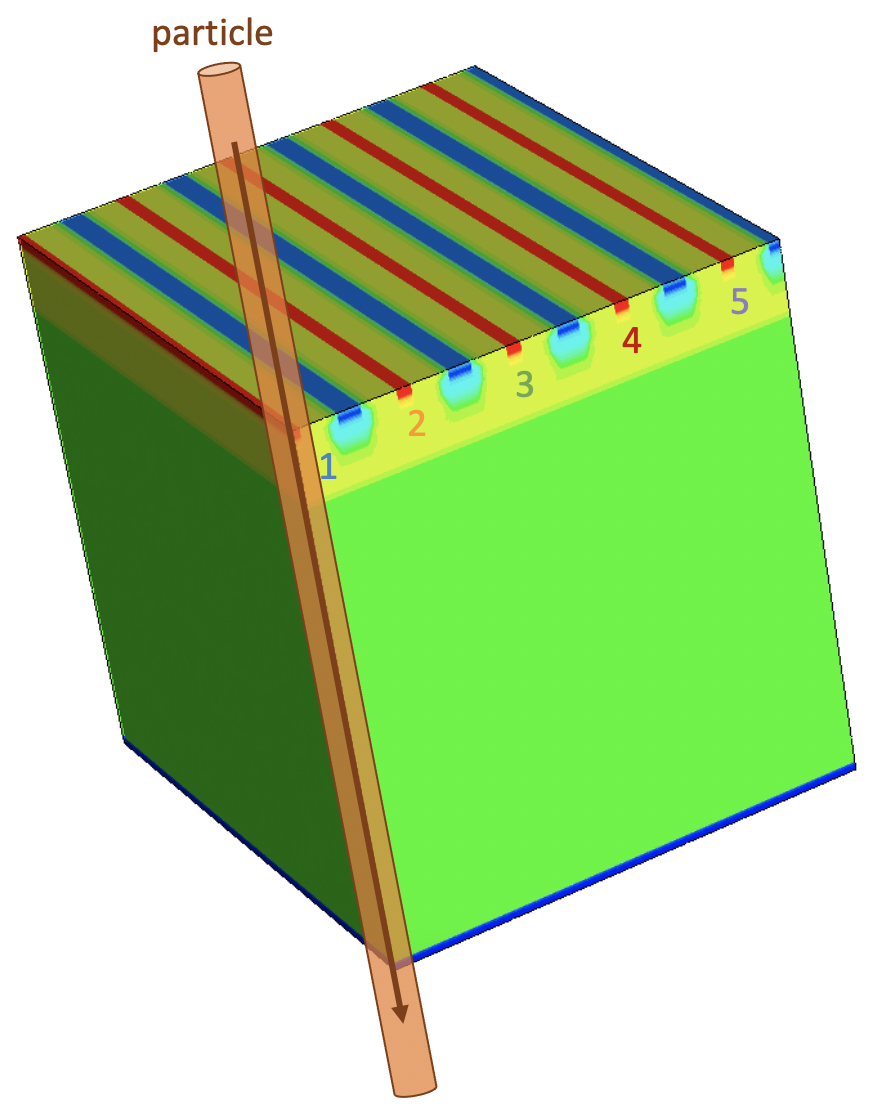}
\caption{Example reduced TCAD domain used in transient simulations (best-case scenario). The microstrips are labelled following the nomenclature illustrated in Figure~\ref{fig:best_worst}. A crossing particle is represented as an orange cylinder hitting the corner of the simulated reduced domain.}
\label{fig:reduced_domain}
\end{figure}

In order to save computational time, a reduced TCAD simulation domain that employs the symmetries was used. This reduced domain corresponds to a quarter of the full domain, with the particle incident in the corner of the domain instead of in the centre. An example for the best-case scenario is shown in Figure~\ref{fig:reduced_domain}. The collected charge and current signals were then scaled to reproduce the full domain case, which includes nine or ten 100$\mu$m long microstrips in the best-case and worst-case scenario respectively (Figure~\ref{fig:best_worst}). These numbers and size of strips guarantee that that the amount of deposited charge reaching the borders of the simulation domain is negligible. The correctness of this strategy was verified and confirmed by comparing the results of a simulation with a quarter domain and of a simulation with full domain.

\vspace{-0.2cm}

\begin{figure}[H]
\centering
\includegraphics[width=14.5 cm]{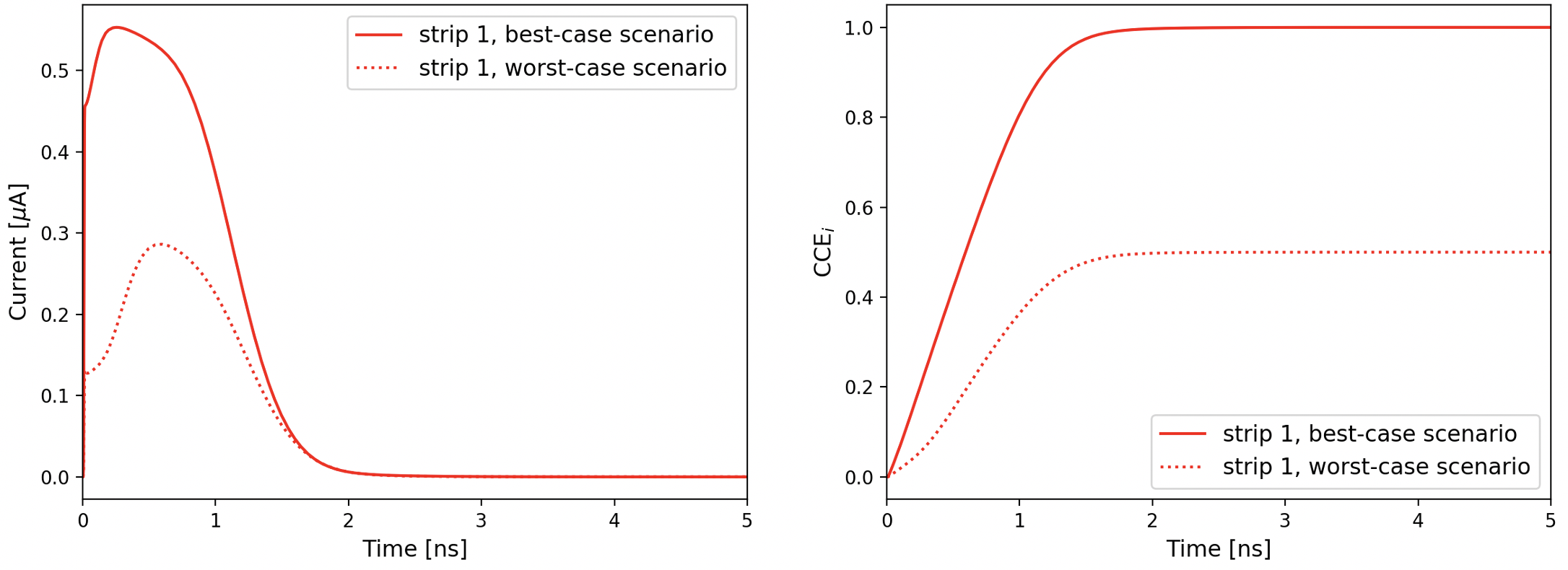}
\caption{Simulated current signals (left) and corresponding charge collection efficiency CCE$_i$ (right) in the best-case and worst-case scenarios for strip number 1 in an example 50\,$\mu$m thick microstrip domain. A particle track with an LET of $1.28 \cdot 10^{-5}$\,pC/$\mu$m was simulated.}
\label{fig:example_signal}
\end{figure}

An example of current signals $I_{nwell}(t)$ measured at the microstrip sensing nodes when a particle crosses the microstrip domain is shown in Figure~\ref{fig:example_signal} (left). We defined as charge collection efficiency for the i-th strip (CCE$_i$) the integral of the current signal $I_{nwell,i}(t)$ extracted from the i-th strip and normalised at the total charge $Q_{tot}$ deposited in the sensor by the particle, according to the formula

\begin{equation}
CCE_i(t) = \frac{\int_{0}^{t} I_{nwell, i}(t’) \,dt’}{Q_{tot}} = \frac{\int_{0}^{t} I_{nwell, i}(t’) \,dt’}{LET \cdot d_{Si}}
\end{equation}

where $d_{Si}$ is the sensor thickness. The total charge collection efficiency CCE for the whole simulated domain is defined as

\begin{equation}
    CCE(t) = \sum_{i=1}^{N_{strips}} CCE_{i}(t)
\end{equation}

where $N_{strips}$ is the total number of strips in the simulated domain. The total CCE at the end of the charge collection process (i.e. at $t = t_{end} = 30$\,ns, which was observed to be large enough for complete charge collection) has to be equal to 100\% in the absence of recombination:

\begin{equation}
CCE (t=t_{max}) = 100\%
\end{equation}

The CCE$_i$ as a function of time is shown in Figure~\ref{fig:example_signal}, on the right, for strip number 1. The times needed for collecting the 95\% and 99\% of the total deposited charge were evaluated and referred to as $t_{95}$ and $t_{99}$, respectively. These values were compared for different design options and used to select the layouts of the fastest sensor flavours.\\

The spatial mesh of the transient simulations was forced to be finer around the particle trajectory to more accurately simulate the charge deposition and the drift of electrons and holes from their generation points along the particle track towards the electrodes. Additionally, the time step of the transient simulations was fine tuned to guarantee the necessary accuracy while keeping the computational time requirement economical. We observed that these adjustments prevented the simulations from giving unphysical results.

\subsection{Determination of the LET for heavy nuclei}

\begin{figure}[H]
\centering
\includegraphics[width=8.7 cm]{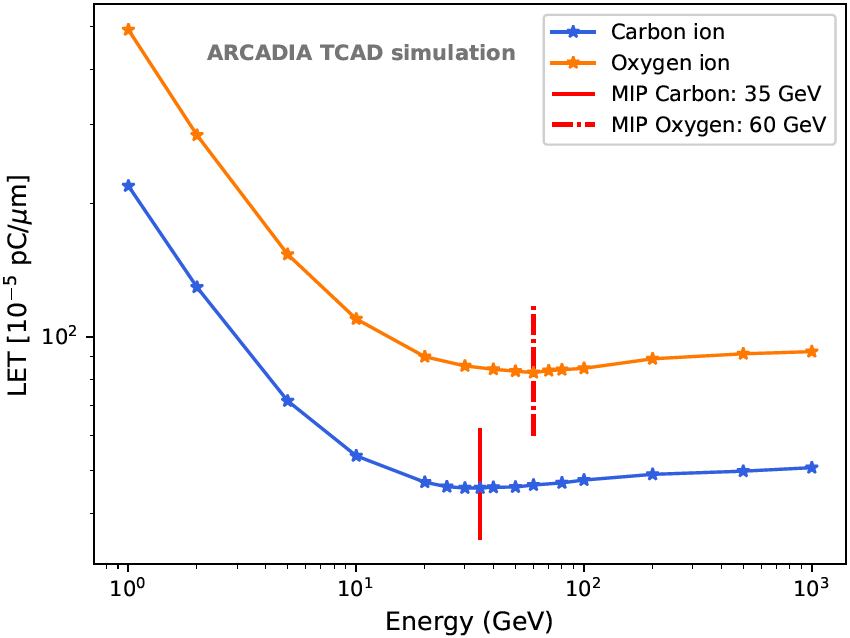}
\caption{Dependence of the LET on the energy of carbon ions (C$^{12+}$, blue) and oxygen ions (O$^{16+}$, orange) incident on 50\,$\mu$m~thick silicon. The LET values were evaluated through Geant4 simulations. The red vertical lines indicate the minimum ionisation energies for the two particle species.}
\label{fig:C_O_LET}
\end{figure}

Since MAMS are an interesting candidate for tracking detectors in space applications, the charge collection was studied not only for minimum ionising particles (MIPs) but also for heavy nuclei of interest for in-orbit astroparticle experiments. The LET values of carbon and oxygen ions were studied in \textsc{Geant4} (version 10.6 patch 01) simulations, and the typically used~\cite{olive2014pdg} LET of 80 electron-hole (e-h) pairs per $\mu$m, or $1.28 \cdot 10^{-5}$\,pC per $\mu$m in silicon for MIPs could be reproduced. The \textsc{Geant4} simulation setup included a 50\,$\mu$m thick silicon layer immersed in air and with a transverse size of 1\,$\times$\,1\,cm$^{2}$. The particle gun was positioned 15\,cm in front of the centre of the silicon layer. The \textit{G4EmPenelopePhysics} physics list was used to model the electromagnetic processes and the necessary precision on the energy deposited within the silicon was achieved with a maximum step size of 1\,$\mu$m~\cite{muonsilicon2020}. The LETs for carbon (C$^{12+}$) and oxygen (O$^{16+}$) ions at their minimum ionisation were computed from their most probable energy loss (i.e. the most probable value of the straggling or Landau functions~\cite{tanabashi2018pdg, bichsel2006method}). Figure~\ref{fig:C_O_LET} shows the LET as a function of the particle energy obtained for C and O ions traversing 50\,$\mu$m of silicon. The energies $E_{min}$ at which C and O ions are at the minimum of ionisation were found to be 35\,GeV and 60\,GeV, respectively. The corresponding LETs are $45.6 \cdot 10^{-5}$\,pC/$\mu$m and $83.0 \cdot 10^{-5}$\,pC/$\mu$m, which results in 36 and 65 times the MIP value. This is consistent with the expected scaling from the Bethe-Bloch formula.

We were especially interested in studying the charge sharing among the microstrips surrounding the particle impact point and the charge collection time at different LETs. This will be reported and discussed in Section~\ref{section:results}.

\section{Results and discussions}
\label{section:results}

In this section, the results of the TCAD simulation campaign will be presented. Their implications will be discussed and their connections to the design objectives will be highlighted. As mentioned in Section~\ref{section:introduction}, the main targets of the FD-MAMS design were the following.
\begin{enumerate}[leftmargin=*,labelsep=4.9mm]
\item To enhance the spatial resolution. A very fine pitch of 10\,$\mu$m was chosen to reach this goal. Intrinsic spatial resolution in case of digital readout would be equal to $\frac{pitch}{\sqrt{12}} = \frac{10\,\mu m}{\sqrt{12}} \simeq 2.9$\,$\mu$m, which can be further improved thanks to charge sharing and with an analog readout.
\item To minimise the sensor capacitance $C_{sens}$ at $V_{back} = V_{pt}$. A low sensor capacitance is particularly important to keep low electronic noise and, consequently, to maximise the SNR.
\item To obtain fast and uniform charge collection, irrespective of the particle incidence position. This will enhance the sensor timing capabilities and will reduce the dead-time between successive particle detections.
\end{enumerate}

For reasons of space available for MAMS in the first ARCADIA engineering run, only a few sensor flavours could be included. Hence, a simulation campaign was needed to identify the best performing sensor layouts. The deep p-well, when present, was kept the same size as the p-well. The expression "p-well and deep p-well" will be contracted and referred to as "(deep) p-well". In the legends of the figures, the abbreviation "dpw" will be used for deep p-well.

\subsection{SiO$_2$ layer and surface damage}
\label{subsec:result_sio2_surf_damage}

\begin{figure}[H]
\centering
\includegraphics[width=8.8 cm]{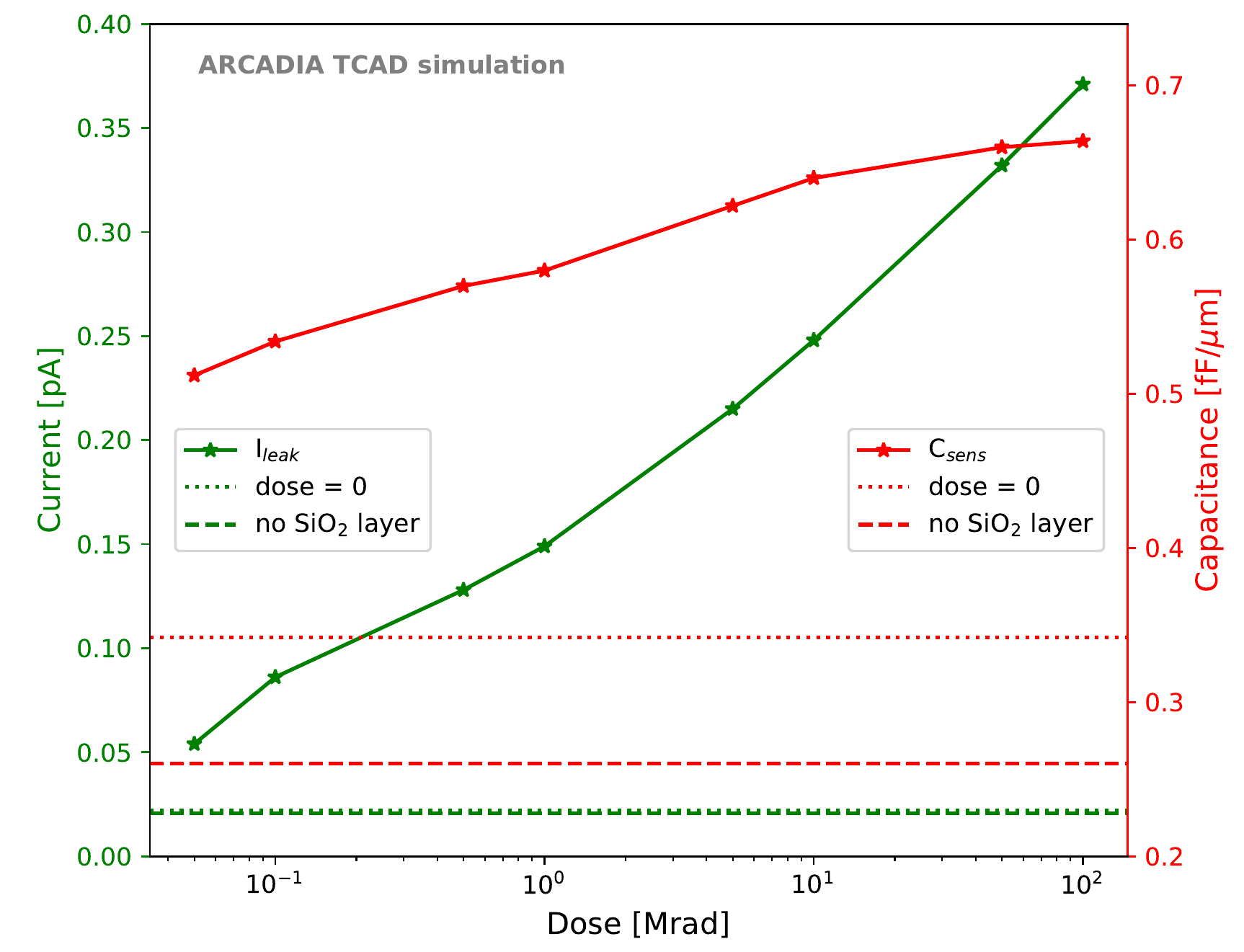}
\caption{Leakage current $I_{leak}$ (green) and sensor capacitance $C_{sens}$ (red) as a function of the total ionising dose for a 50\,$\mu$m thick microstrip sensor. The values obtained in simulations with and without the SiO$_2$ layer in the absence of irradiation are shown as horizontal lines and referred to as "dose~=~0" and "no~SiO$_2$~layer" respectively.}
\label{fig:leak_csens_dose}
\end{figure}

\begin{figure}[H]
\centering
\includegraphics[width=9 cm]{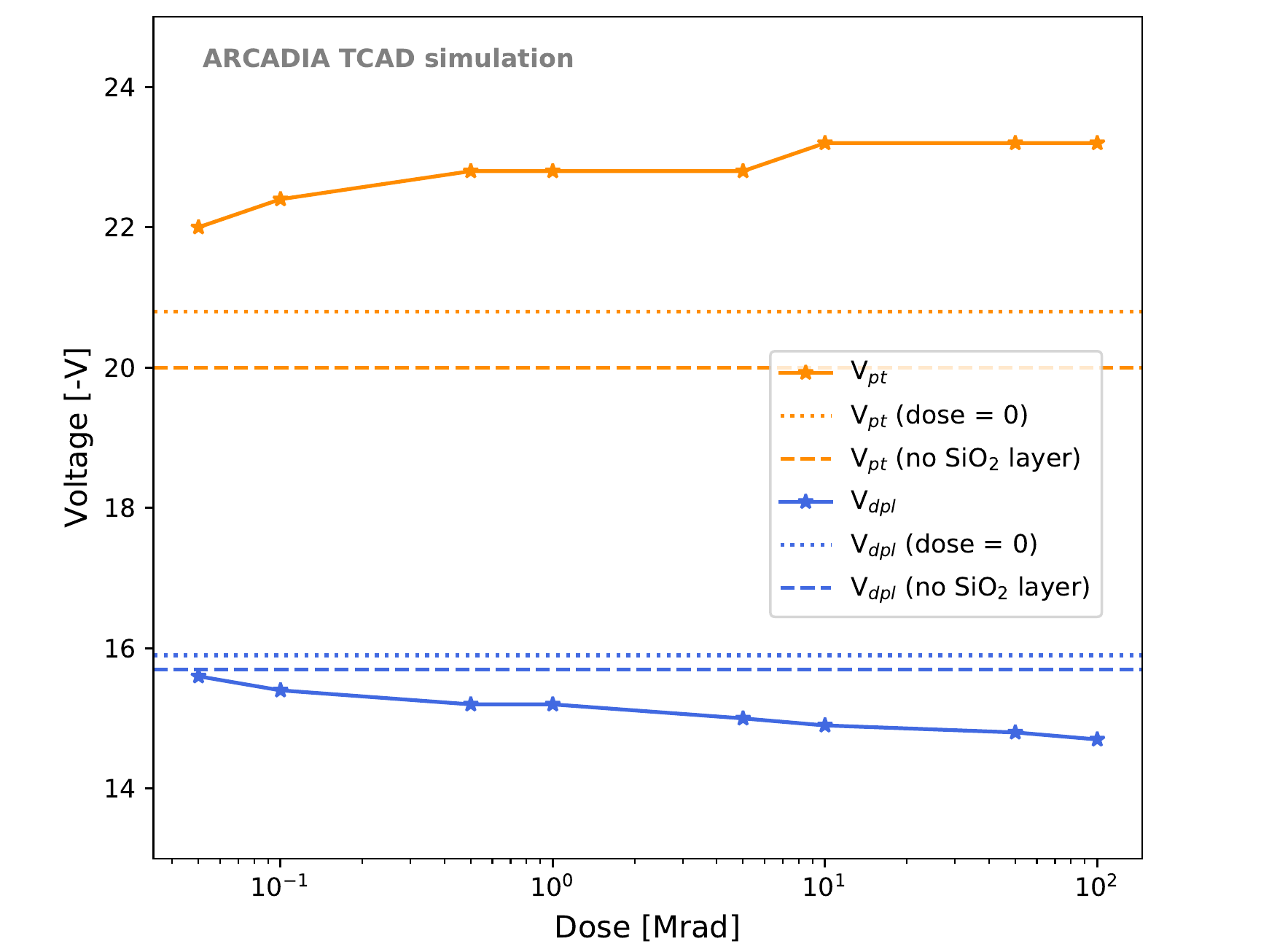}
\caption{Depletion voltage $V_{dpl}$ (blue) and punch-through voltage $V_{pt}$ (orange) as a function of the total ionising dose for a 50\,$\mu$m thick microstrip sensor. The values obtained in simulations with and without the silicon dioxide layer in the absence of irradiation are shown as horizontal lines.}
\label{fig:vdpl_vpt_dose}
\end{figure}

A first group of TCAD simulation studies was aimed at investigating the effects of the SiO$_2$ layer and of surface TID damage on the FD-MAMS characteristics. The model that we employed was presented in Paragraph~\ref{subsubsec:sio2_radiation}. As can be seen from Figure~\ref{fig:leak_csens_dose}, for one of the selected 50\,$\mu$m thick microstrip layouts, the inclusion of the SiO$_2$ layer with a minimum concentration of traps and oxide charges ($dose = 0$) determines a small increase of about 5\% in the leakage current $I_{leak}$ from 20.8\,fA to 22.0\,fA. The sensor capacitance $C_{sens}$ is strongly affected by the inclusion of the SiO$_2$ layer, as it increases by 31\% from 0.26\,fF/$\mu$m to 0.34\,fF/$\mu$m. Both $I_{leak}$ and $C_{sens}$ are found to rise with increasing dose. The minimum dose that we considered is 50~krad, as the model is not validated for lower doses~\cite{passeri2019tcad}. Figure~\ref{fig:vdpl_vpt_dose}, instead, shows the effect of the SiO$_2$ layer and of the TID on $V_{dpl}$ and on $V_{pt}$. The effect of the dose on these two values is smaller than in the case of $I_{leak}$ and $C_{sens}$. Furthermore, $V_{dpl}$ and $V_{pt}$ are influenced by the dose in opposite directions, which results in a slight increase in the operating range $\Delta$V$_{op}$ with increasing dose.

\subsubsection{Effect on sensor capacitance}
\label{subsubsec:result_effect_on_csens}

\begin{figure}[H]
\centering
\includegraphics[width=7.8 cm]{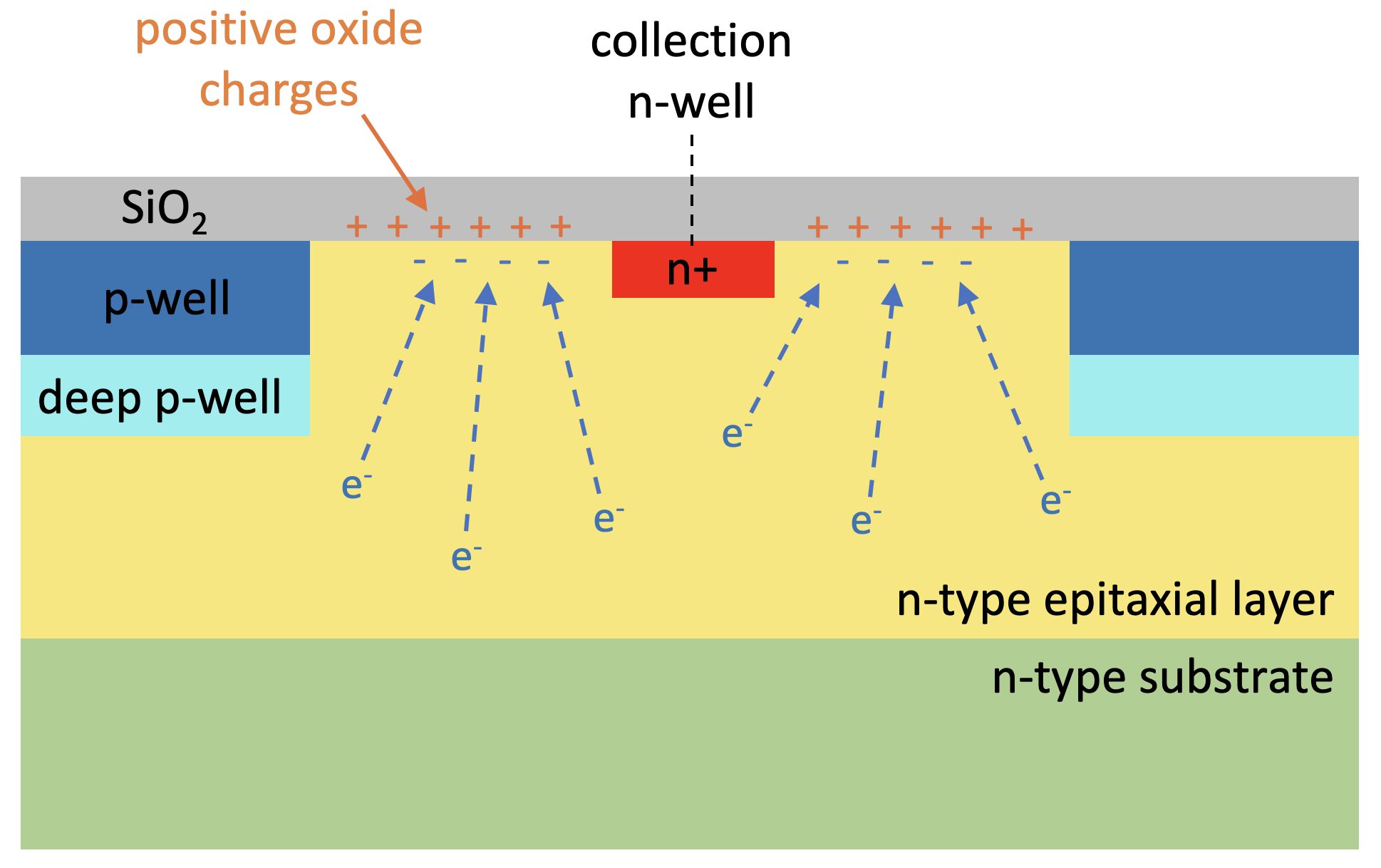}
\caption{Schematic illustration of the electron accumulation in the gap between the collection n-well and the surrounding p-wells due to the positive oxide charges introduced at the Si-SiO$_2$ interface.}
\label{fig:sio2_electron_attraction}
\end{figure}

\begin{figure}[H]
\centering
\includegraphics[width=\textwidth]{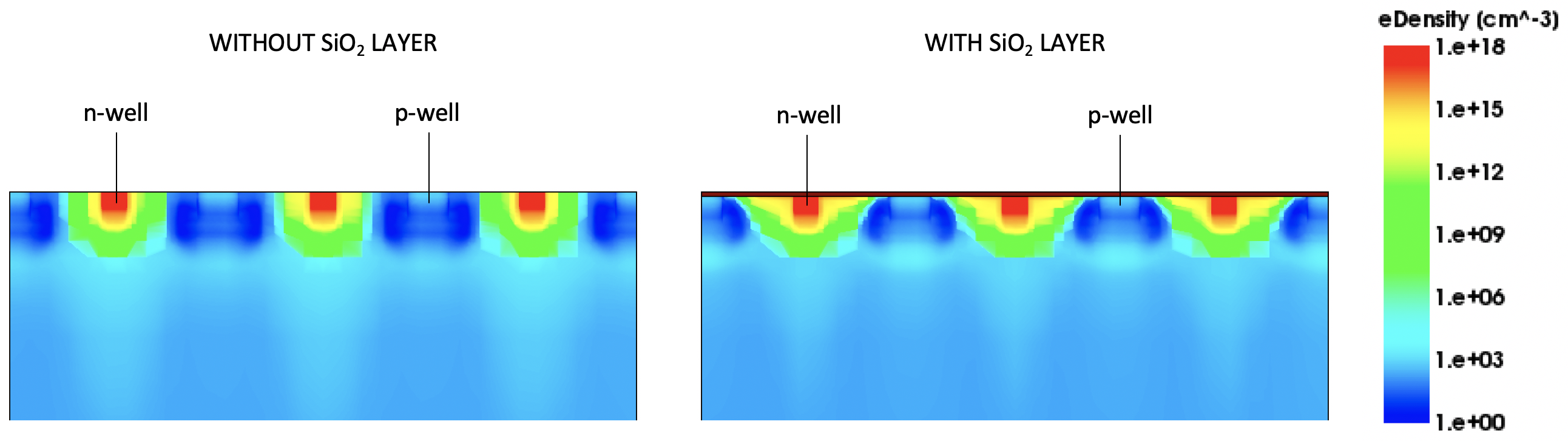}
\caption{Electron density in an example microstrip simulation domain without (left) and with (right) the SiO$_2$ layer on top of the sensors.}
\label{fig:nwell_extension}
\end{figure}

The reason for the significant capacitance increase even after the simple inclusion of the SiO$_2$ layer was found to be due to the introduction of positive oxide charges at the Si-SiO$_2$ interface~\cite{neubueser2020sensor}. In fact, the model that we adopted foresees a significant positive oxide charge concentration $Q_{ox}$\,=\,6.5\,$\cdot$\,$10^{10}$\,charges/cm$^{-2}$ already at dose~=~0. These positive oxide charges attract free electrons from the n-type silicon epitaxial layer towards the Si-SiO$_2$ interface in the gap and determine an increase in the electron concentration around the heavily n-doped collection well, as illustrated in Figures~\ref{fig:sio2_electron_attraction} and~\ref{fig:nwell_extension}. This electron accumulation behaves as an extension of the collection n-well.

\vspace{-0.1cm}

\vspace{-0.2cm}

\subsection{Capacitance minimisation}
\label{subsec:csens_minimisation}

\begin{figure}[H]
\centering
\includegraphics[width=9.2 cm]{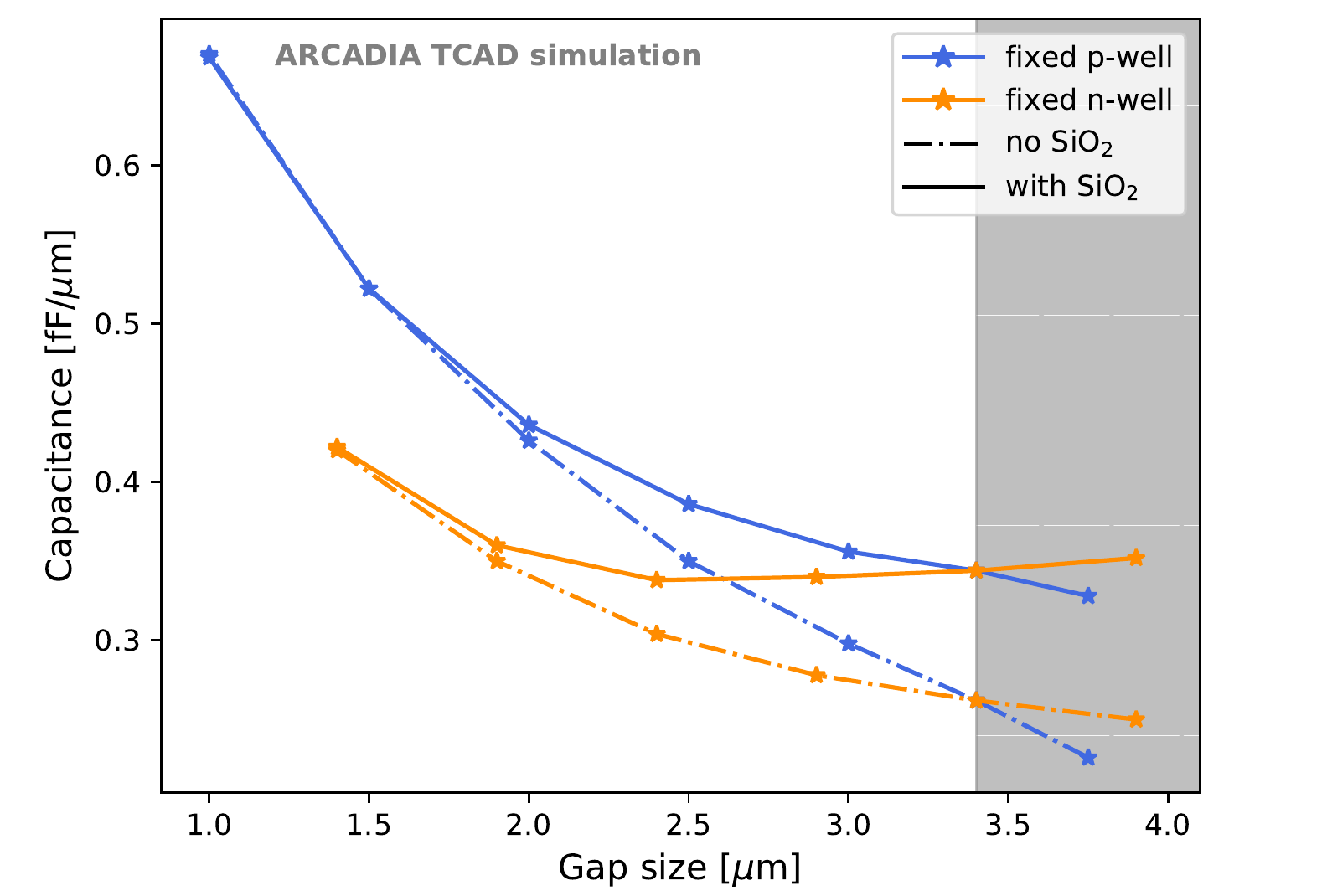}
\caption{$C_{sens}$ as a function of the gap size for different sensor layout configuration. The vertical grey band is the forbidden region due to fabrication constraints; its leftmost limit is still permitted.}
\label{fig:c_at_vpt_vs_gap}
\end{figure}

The sizes of both the collection n-well and of the gap were found to contribute to $C_{sens}$. Therefore, both n-well and (deep) p-well sizes were adjusted to find the optimal layout for $C_{sens}$ minimisation. It was observed the inclusion of the SiO$_2$ layer influences $C_{sens}$ in different ways for different gap sizes. Hence, $C_{sens}$ with and without the SiO$_2$ layer was evaluated.
Figure~\ref{fig:c_at_vpt_vs_gap} shows the trend of $C_{sens}$ as a function of the gap size for 50\,$\mu$m thick microstrips. The different sensor thicknesses considered (50, 100 and 300\,$\mu$m) were found not to influence the sensor capacitance. Both the case with fixed minimum-size n-well and variable (deep) p-well (blue curves) and the case with fixed minimum-size (deep) p-well and variable n-well (orange curve) were studied. The dash-dotted lines refer to simulations without the surface SiO$_2$ layer, whereas solid lines to the case with SiO$_2$ layer included with minimal oxide charge and trap concentration.

The reason for which smaller gaps with fixed n-wells could not be investigated is referred to as \textit{channel choking}, a condition that inhibits sensor operation; this condition is explained in Section~\ref{subsec:operating_voltages}. The vertical grey band in Figure~\ref{fig:c_at_vpt_vs_gap} and in the following ones corresponds to the forbidden region due to the constraints on n-well and (deep) p-well minimum sizes imposed by the fabrication process. The leftmost limit of the grey band is still permitted.

Variations of n-well and of (deep) p-well size do not lead to the same $C_{sens}$ for the same gap size. A fixed n-well size with SiO$_2$ layer included shows a trend that is not monotonic, but has a minimum at slightly less than 0.34\,fF/$\mu$m. This effect is caused by the electron accumulation in the gap at the Si-SiO$_2$ interface. However, the difference in $C_{sens}$ between the minimum-capacitance option and the sensor layout at the edge of the forbidden region is lower than 2\%. There was, as expected, no benefit found from having large n-wells. The sensor capacitance increases with the n-well size, as can be seen from the blue curve in Figure~\ref{fig:c_at_vpt_vs_gap}. Therefore, we chose the best layout for minimum $C_{sens}$ to have the smallest possible n-well size and sufficiently small (deep) p-well.

\begin{figure}[H]
\centering
\includegraphics[width=9.2 cm]{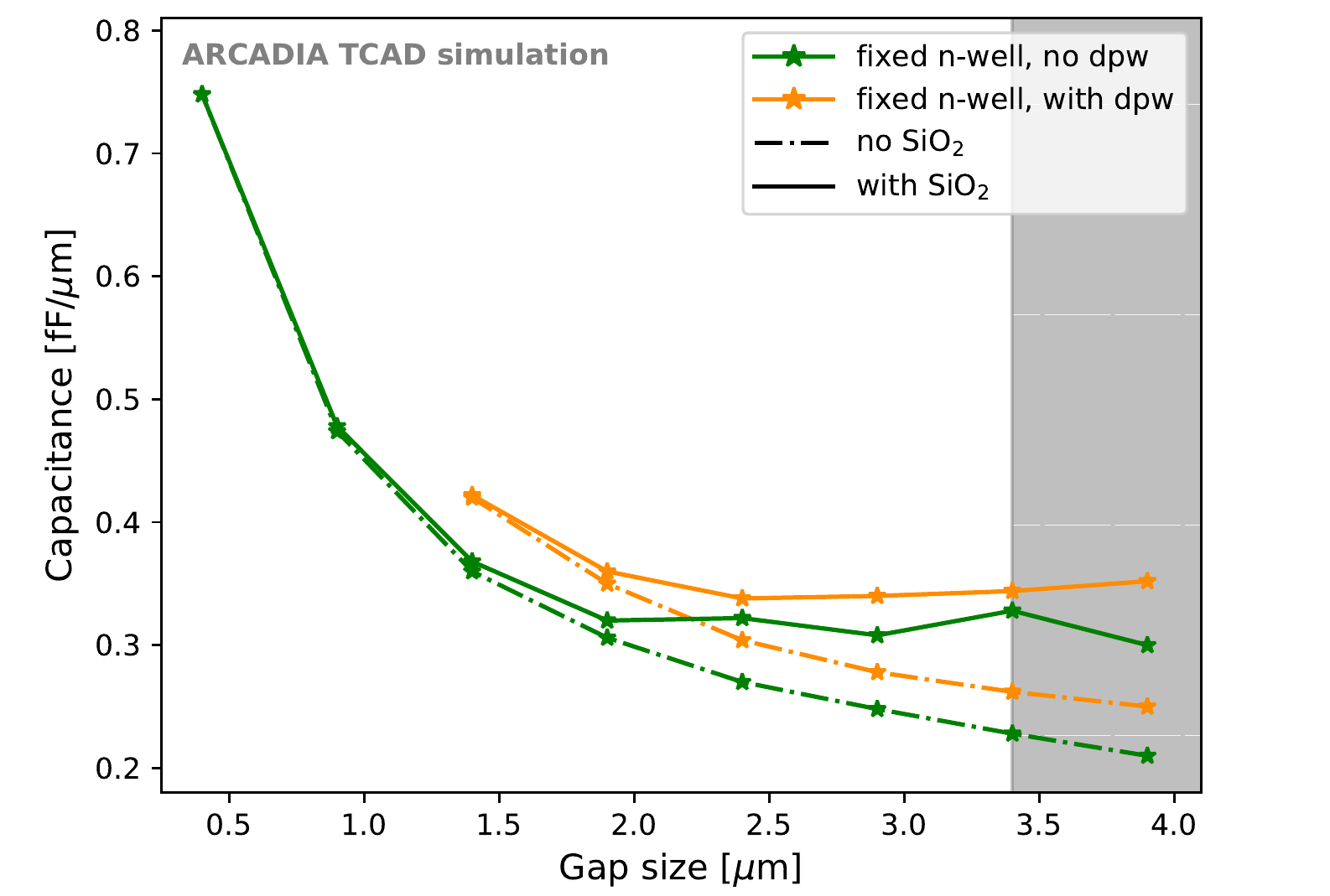}
\caption{Sensor capacitance $C_{sens}$ as a function of the gap size for different sensor layout configurations with and without the deep p-well.}
\label{fig:c_at_vpt_vs_gap_no_dpw}
\end{figure}

Figure~\ref{fig:c_at_vpt_vs_gap_no_dpw} compares the sensor capacitance for layouts with deep p-well (orange curve) and without deep p-well (green curve). All the sensor flavours feature the minimum n-well size permitted by the fabrication process. On the one hand, removing the deep p-well could help in further reducing the sensor capacitance. On the other hand, this choice would strongly affect the sensor bias voltage operating range, as discussed in the following section.

\subsection{Reference and operating voltages}
\label{subsec:operating_voltages}

\begin{figure}[H]
\centering
\includegraphics[width=9.2 cm]{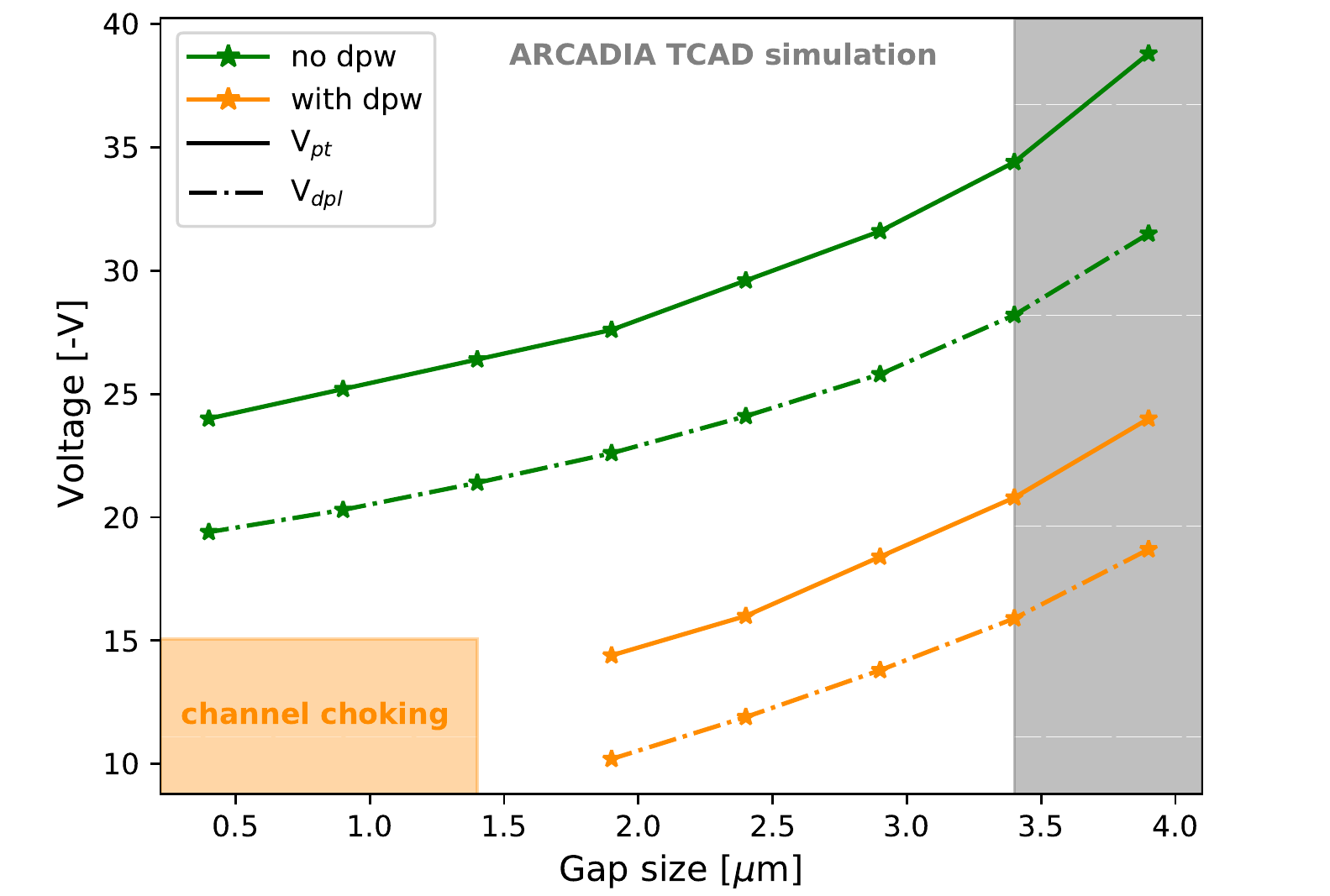}
\caption{Sensor depletion voltage $V_{dpl}$ and punch-through voltage $V_{pt}$ as a function of the gap size for different sensor layout configurations. The orange region indicates the forbidden region due to the observed \textit{channel choking}.}
\label{fig:vdpl_vpt_vs_d_pw_size_choking}
\end{figure}

We found the influence of the n-well size on the operating voltages to be negligible compared to the effect of the (deep) p-well size. Therefore, for the sake of capacitance minimisation, we fixed the n-well size at the smallest possible value. With this assumption, Figure~\ref{fig:vdpl_vpt_vs_d_pw_size_choking} presents the effect of the (deep) p-well size effect on $V_{dpl}$ and on $V_{pt}$ for 50\,$\mu$m thick sensors. Both the cases with (orange curves) and without deep p-well (green curves) were considered and compared. The voltage values are reported for the case of dose\,=\,0.

In all the layouts considered in Figure~\ref{fig:vdpl_vpt_vs_d_pw_size_choking}, the onset of the punch through happens at voltages sufficiently larger than the depletion voltage. Outside of the forbidden region (grey band), the operating range $\Delta V_{op}$ is always between 4.2\,V and 6.2\,V, or between the 23\% and the 41\% of $V_{dpl}$. This is a sufficiently large operating range for safe sensor operation, even in the hypothesis of possible doping inhomogeneities among adjacent microstrips or slight deviations from the doping design values. Similar observations on $\Delta V_{op}$ have been made for 100\,$\mu$m thick and 300\,$\mu$m thick sensors.

As a general trend, it can be observed in Figure~\ref{fig:vdpl_vpt_vs_d_pw_size_choking} that smaller (deep) p-wells result in larger $V_{dpl}$ and $V_{pt}$. This can be interpreted as follows. Large p-doped surfaces below the (deep) p-wells create wider pn-junctions with the n-doped epitaxial layer, thus facilitating the depletion of the underlying epitaxial layer at lower voltages. On the other hand, large (deep) p-wells also lower the potential barrier that prevents the direct flow of holes towards the substrate. This results in the earlier onset of the punch through hole current between the (deep) p-wells and the backside p+ region.
Sensors without the deep p-well showed higher reference voltages. In fact, the presence of a deep p-well reduces the epitaxial layer thickness below the p-wells, thus requiring a lower voltage to achieve both full depletion and the onset of punch-through currents.

Finally, for sensors with too large deep p-well, a phenomenon that we defined as \textit{channel choking} was observed. This consists in the closure of the conductive channel below the collection n-well due to the lateral merging of the closely adjacent depletion regions formed at the junctions between the deep p-wells and the n-epitaxial layer. In this situation, in the simulations performed to extract $V_{dpl}$, no current flows among the n-wells at low values of $V_{back}$, even though the space charge region of the backside junction has not reached the surface yet. In this condition, the $I_{nwell,unbalanced}$ curve, that corresponds to the orange curve shown in Figure~\ref{fig:iv_cv}, appears flat and no $V_{dpl}$ can be extracted. This means that the n-wells are already isolated from one another at $V_{back} = 0$\,V and that the process of charge collection, which generates the current $I_{nwell}$ measured at the sensing node, is inhibited by the strong potential barrier present below the n-wells. No channel choking was observed for sensor layouts without the deep p-well.\\

\begin{figure}[H]
\centering
\includegraphics[width=\textwidth]{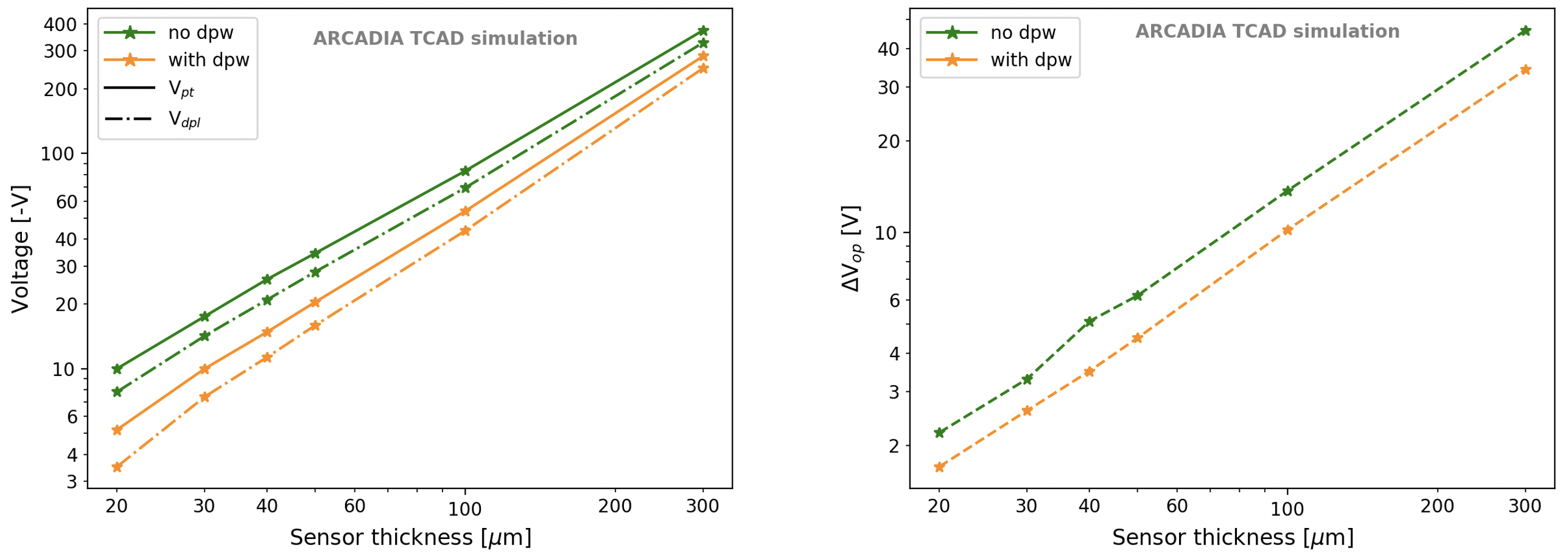}
\caption{Dependence of $V_{dpl}$ and $V_{pt}$ (left) and of the operating voltage range $\Delta V_{op}$ (right) on the sensor thickness.}
\label{fig:vdpl_vpt_vs_thk}
\end{figure}

For completeness, Figure~\ref{fig:vdpl_vpt_vs_thk} (left) illustrates the dependence of $V_{pt}$ and $V_{dpl}$ on the sensor thickness for the sensor layout with minimum sizes for the n-well and for the (deep) p-well. The trend is linear over a wide range of thicknesses, both with and without the deep p-well. Also the operating voltage $\Delta V_{op} = V_{pt} - V_{dpl}$ linearly increases with the sensor thickness, as shown in Figure~\ref{fig:vdpl_vpt_vs_thk} (right). The sensor thickness investigated was extended down to 20\,$\mu$m, well below the smallest thickness (i.e. 50\,$\mu$m) of the sensors that will be produced in the first ARCADIA engineering run. The reason for this will become clear in Paragraph~\ref{subsubsec:timing_enhance}, as the study of very thin sensors was functional for enhancing the speed of the charge collection process and, consequently, for improving the sensor timing performance.

\vspace{-0.1cm}

The voltage $V_{pd}$ at which the power density is 0.1\,mW/cm$^2$ was found to be about 4-5\,V above $V_{pt}$ for 50\,$\mu$m thick microstrips, 7-8\,V for 100\,$\mu$m thick microstrips and 18-20\,V for 300\,$\mu$m thick microstrips when the deep p-well was included.

\subsection{Effects of V$_{nwell}$}
\label{subsec:vn_effects}

\begin{figure}[H]
\centering
\includegraphics[width=\textwidth]{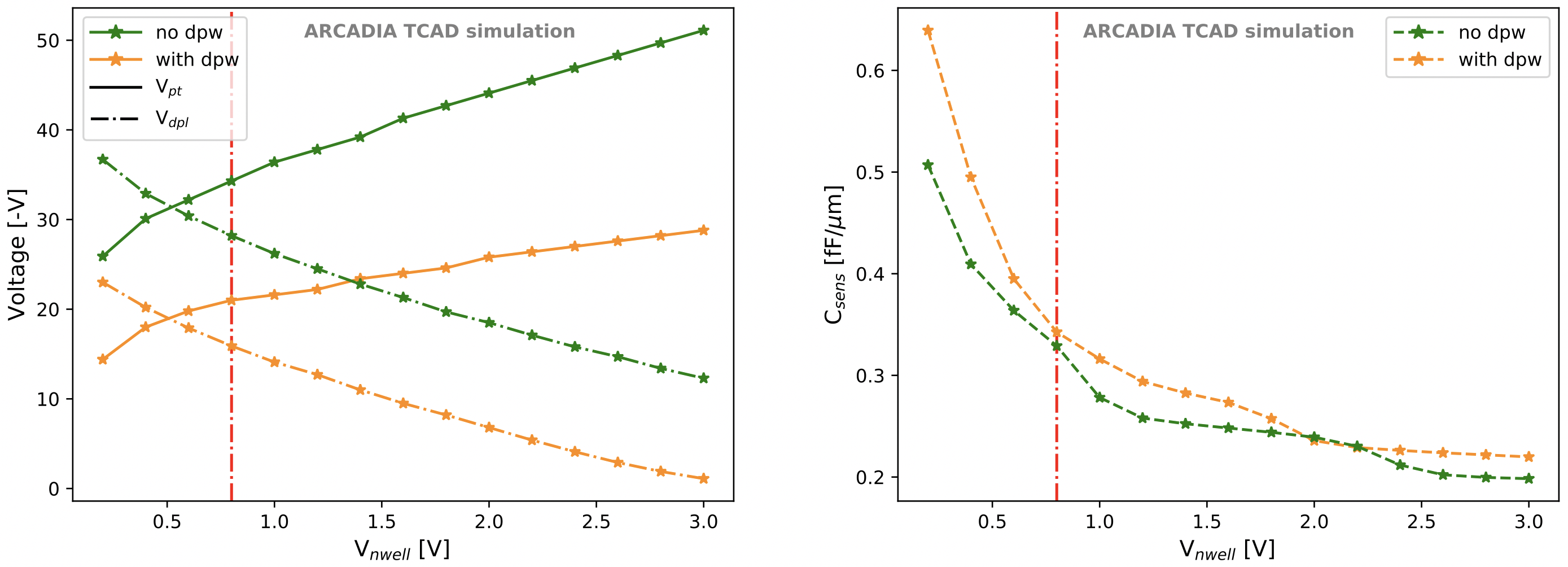}
\caption{$V_{dpl}$ and $V_{pt}$ (left) and sensor capacitance (right) as a function of $V_{nwell}$.  The vertical red line indicates the default value of $V_{nwell} = 0.8$\,V.}
\label{fig:vdpl_vpt_vs_vnwell}
\end{figure}

The $V_{nwell}$ voltage was varied with the aim of finding possible improvements in the sensor performances. The results are shown in Figure~\ref{fig:vdpl_vpt_vs_vnwell} (left), where the vertical red line indicates the default value of 0.8\,V. A minimum $V_{nwell}$ of about 0.5\,V is necessary to satisfy the condition $|V_{pt}| > |V_{dpl}|$. Moreover, an increase in $V_{nwell}$ has several interesting effects. First of all, it allows the sensor full depletion to be reached at lower (in absolute value) backside voltages. Secondly, it also shifts the onset of the punch through towards larger $|V_{back}|$, thus increasing the operating range $\Delta V_{op}$. Finally, as shown in Figure~\ref{fig:vdpl_vpt_vs_vnwell} (right), larger $V_{nwell}$ implies lower sensor capacitance.

\subsection{Charge collection studies}
\label{subsec:charge_collection}

\begin{figure}[H]
\centering
\includegraphics[width=9.2 cm]{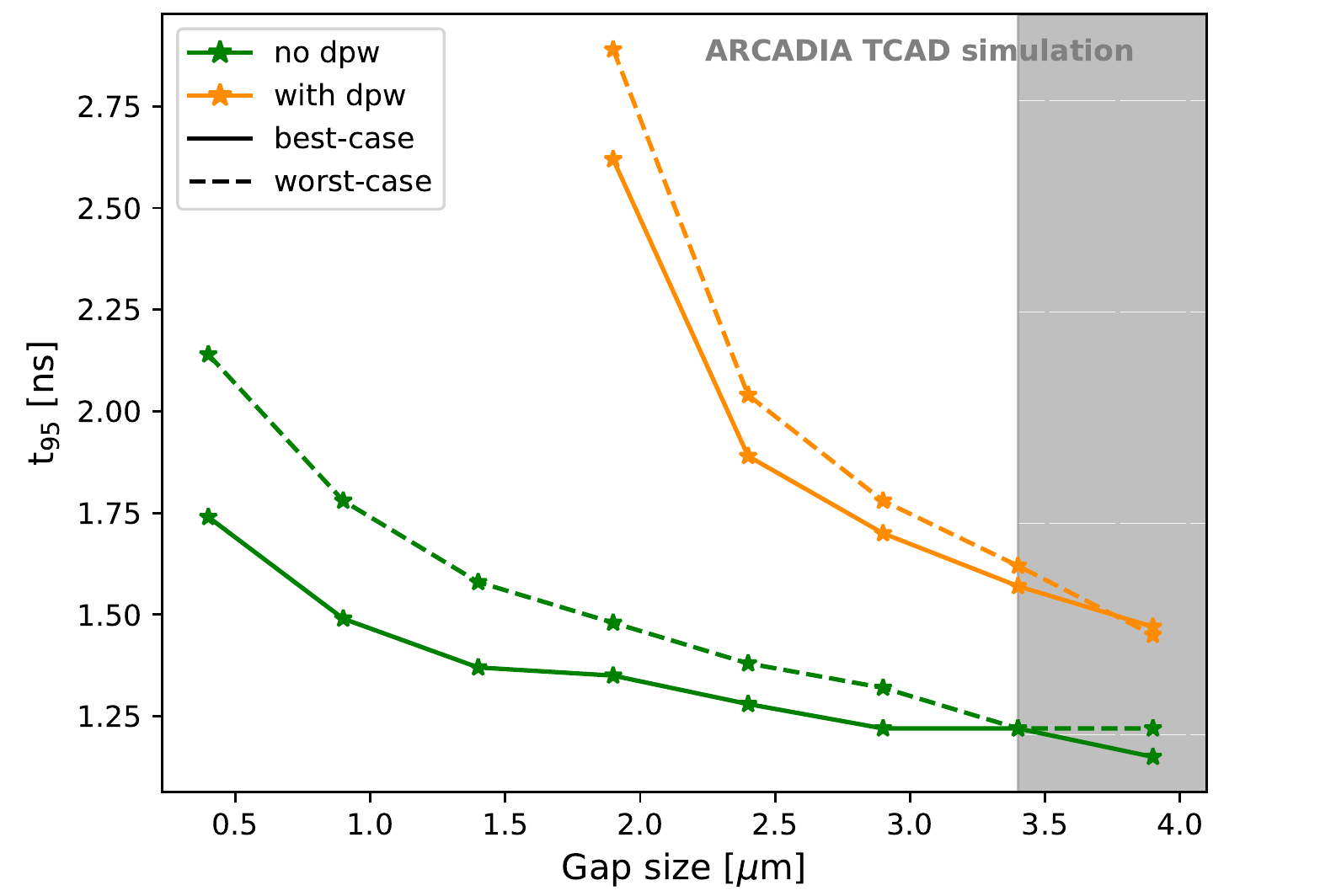}
\caption{$t_{95}$ as a function of the gap size for best-case and worst-case scenarios.}
\label{fig:t95_vs_gap}
\end{figure}

As described in Paragraph~\ref{subsubsec:transient_sim}, TCAD transient simulations were used to study the charge collection dynamics. In order to select the layouts with the optimal performance in terms of fast and uniform charge collection, the effect of the (deep) p-well size on the charge collection time at $V_{back} = V_{pt}$ was evaluated. The time $t_{95}$ needed to collect 95\% of the total charge deposited in the simulated sensor domain is plotted in Figure~\ref{fig:t95_vs_gap} for 50\,$\mu$m thick sensors and LET\,=\,$1.28 \cdot 10^{-5}$\,pC/$\mu$m (1 MIP) as a function of the gap size and with fixed minimum n-well size.

Microstrips with large gaps, hence small (deep) p-wells, are to be preferred for fast charge collection. The reason for this is a higher $|V_{pt}|$, which enables sensor operation at a larger $|V_{back}|$. The consequent stronger electric field in the sensor results in higher charge velocity in the silicon substrate. For the same reason, microstrip sensors without deep p-well revealed a significantly faster charge collection in both the best-case and the worst-case scenario. 
Flavours with small (deep) p-wells also show very uniform charge collection for different particle incidence positions. The difference in $t_{95}$ for the best-case and worst-case scenarios is below 0.1\,ns for the fastest permitted options. This result is also achieved thanks to the fine microstrip pitch of 10\,$\mu$m.
The channel choking, as described in Section~\ref{subsec:operating_voltages}, limits the deep p-well size as the potential barrier below the sensing node slows down the electron collection. This problem, as shown in Figure~\ref{fig:t95_vs_gap}, can be avoided by removing the deep p-well.\\

\begin{figure}[H]
\centering
\includegraphics[width=9.2 cm]{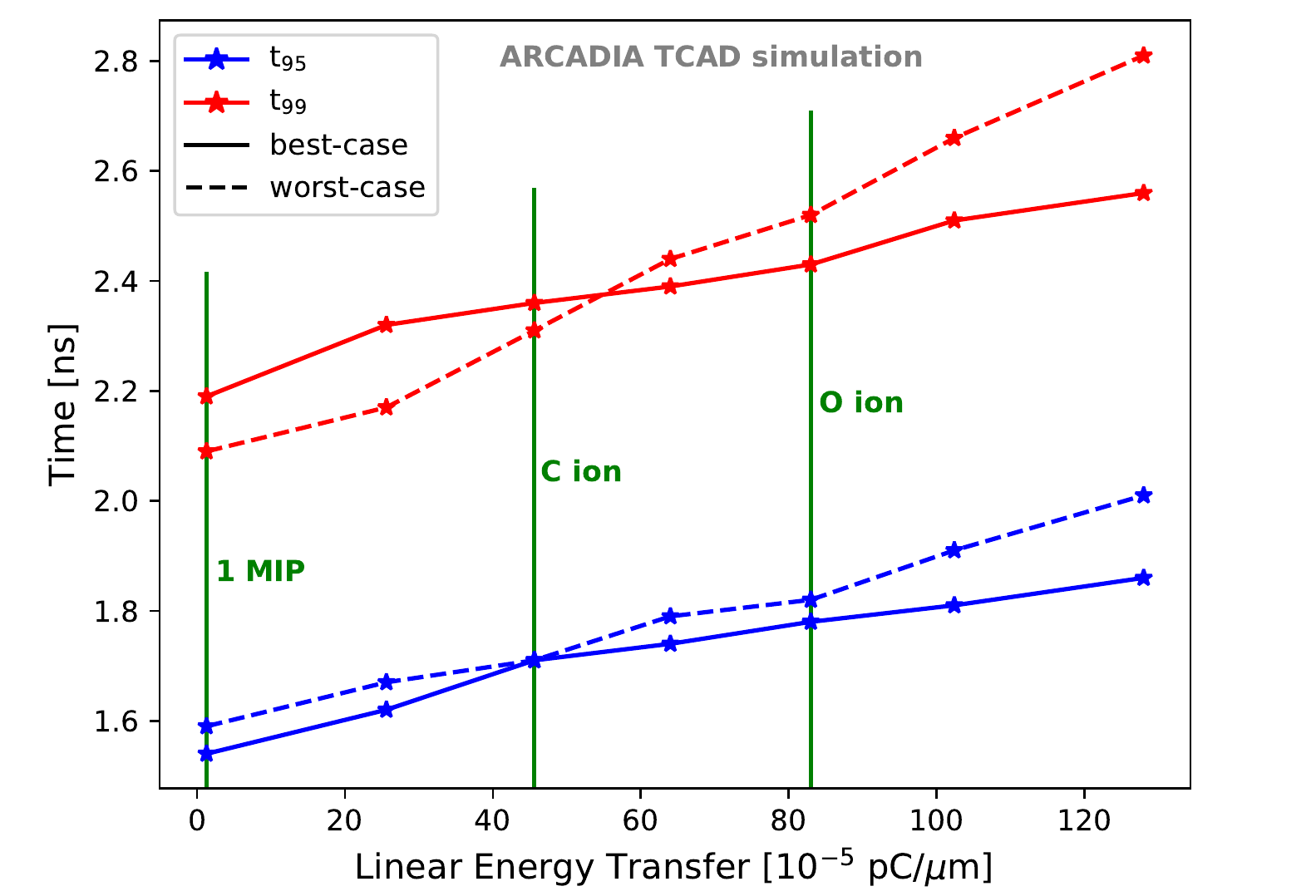}
\caption{$t_{95}$ (blue) and $t_{99}$ (red) as a function of the LET for best-case and worst-case scenarios.}
\label{fig:time_vs_LET}
\end{figure}

Figure~\ref{fig:time_vs_LET} demonstrates that the proposed MAMS guarantee fast sensor response also under heavily ionising particles. The charge collection time is only weakly proportional to the charge deposited by the incident particle within an LET range of [1.28; 128]\,$\cdot$\,10$^{-5}$\,pC/$\mu$m. A 50\,$\mu$m thick sensor was considered in Figure~\ref{fig:time_vs_LET}, and the LET values corresponding to 1\,MIP, carbon (C) ion and oxygen (O) ion at their minimum of ionisation are highlighted as vertical green lines. Moreover, $t_{99}$ is added to show that the time needed for complete charge collection is only slightly larger than $t_{95}$, due to a small fraction of charge collected by the strips adjacent to the central one. However, $t_{95}$ and $t_{99}$ were never found to exceed 2\,ns and 3\,ns respectively in 50\,$\mu$m thick sensors.

\subsubsection{Further enhancements for fast timing performance}
\label{subsubsec:timing_enhance}

As we discussed in Section~\ref{subsec:charge_collection}, the first strategy for improving the timing performance of the proposed microstrip sensors is to remove the deep p-well in order to obtain larger $|V_{pt}|$. However, we also investigated other ways to increase $|V_{pt}|$ and to speed up the charge collection. In particular, as shown in Section~\ref{subsec:vn_effects}, a larger $V_{nwell}$ is capable of shifting the onset of the punch-through current towards larger $|V_{back}|$. Therefore, we explored the effects of $V_{nwell}$ on the charge collection time.\\

In a strip readout system, timing information can be retrieved only from the strips collecting most of the charge (i.e. strip(s) number 1, following the nomenclature of Figure~\ref{fig:best_worst}), as they provide a signal with sufficiently large SNR. Therefore, in order to study the sensor timing performance and after verifying through $t_{95}$ that the total deposited charge is quickly collected in the whole simulation domain, we considered the time $t_{95}^{central}$ needed to collect 95\% of the charge in the central strip(s).

\begin{figure}[H]
\centering
\includegraphics[width=9.2 cm]{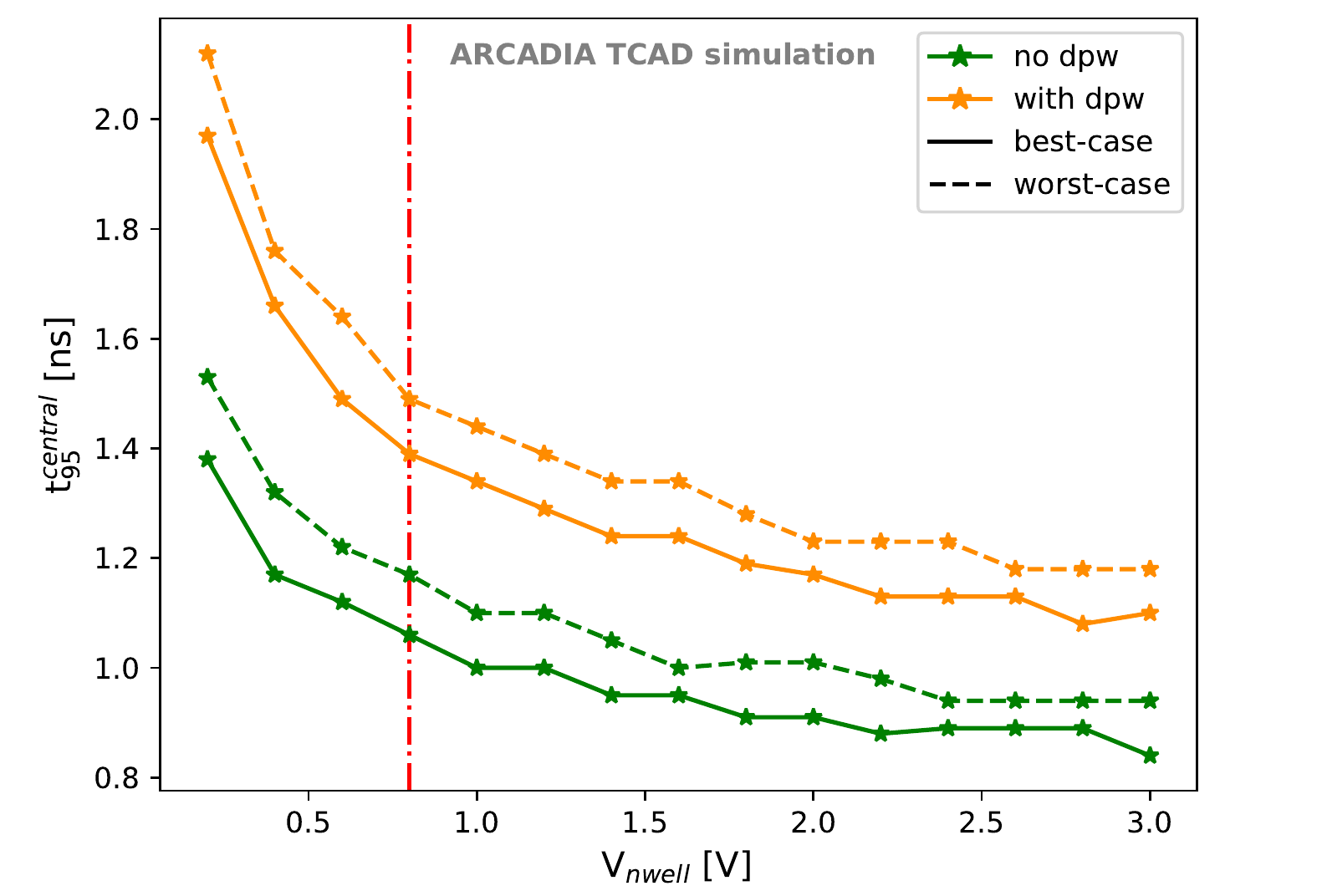}
\caption{$t_{95}^{central}$ as a function of the voltage $V_{nwell}$ applied to the sensing node for best-case and worst-case scenarios. The vertical red line indicates the default value of $V_{nwell} = 0.8$\,V.}
\label{fig:t95_central_vs_vnwell}
\end{figure}

Figure~\ref{fig:t95_central_vs_vnwell} shows the dependence of $t_{95}^{central}$ at $V_{back} = V_{pt}$ and with LET\,=\,$1.28 \cdot 10^{-5}$\,pC/$\mu$m on the voltage applied to the sensing node. A 50\,$\mu$m thick sensor with a layout optimised for fast charge collection was considered. A significant improvement could be reached at larger $V_{nwell}$. For the option without deep p-well and at $V_{nwell} = 3$\,V, $t_{95}^{central}$ is 0.84\,ns in the best-case and 0.94\,ns in the worst-case scenario. If we assume an electron drift saturation velocity of $\sim 1 \cdot 10^7$\,cm/s in silicon at a temperature of 300\,K~\cite{canali1975electron}, the minimum drift time for electrons that have to cover a 50\,$\mu$m distance is 0.5\,ns. This explains the saturation observed in Figure~\ref{fig:t95_central_vs_vnwell} and demonstrates the fast charge collection and the promising timing capabilities of the proposed MAMS.

\begin{figure}[H]
\centering
\includegraphics[width=9.2 cm]{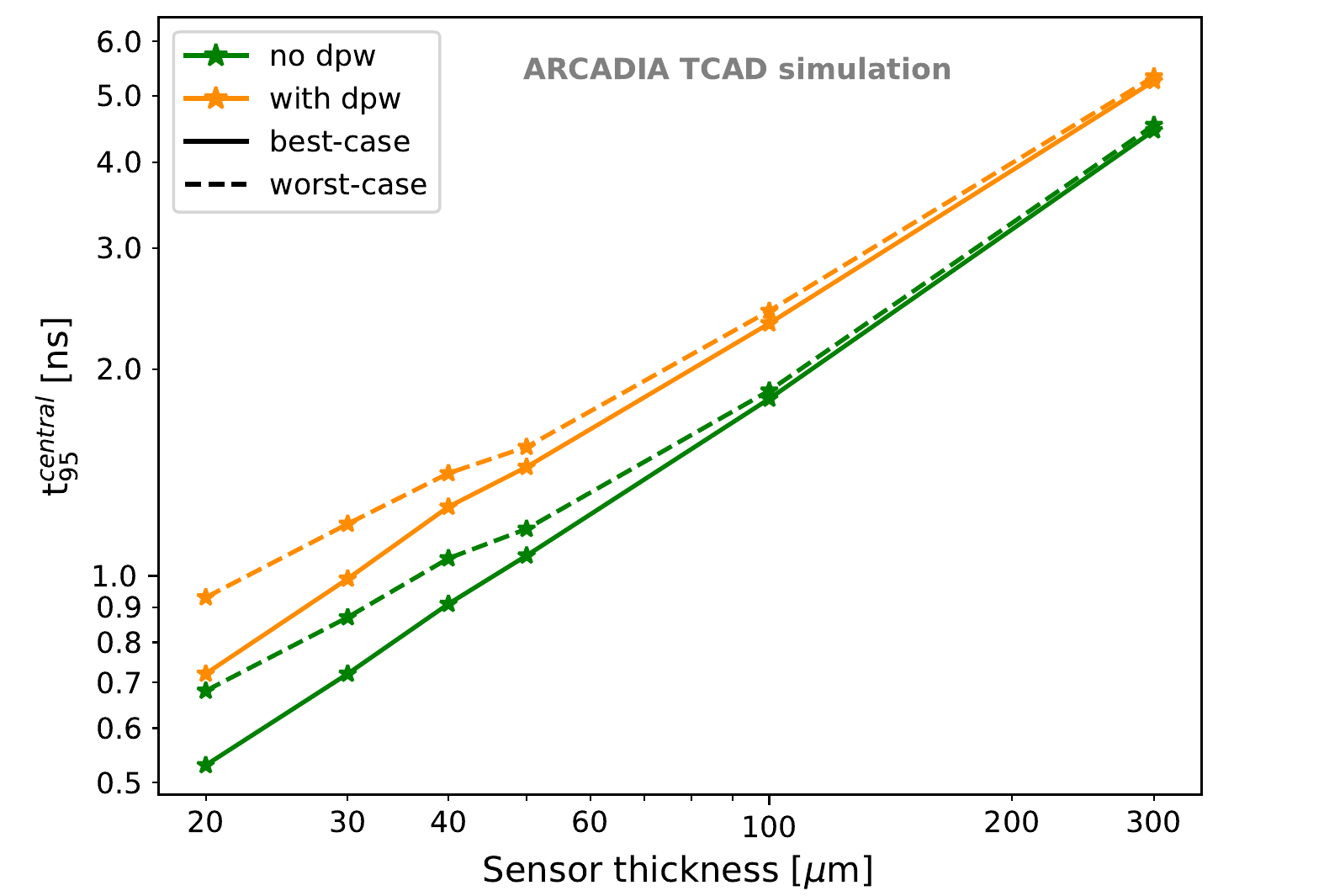}
\caption{$t_{95}^{central}$ as a function of the sensor thickness for best-case and worst-case scenarios.}
\label{fig:t95_central_vs_thk}
\end{figure}

A way to further reduce the collection time is to explore thinner sensors. Figure~\ref{fig:t95_central_vs_thk} demonstrates that the charge collection time $t_{95}^{central}$ is proportional to the sensor thickness. For these simulations, $V_{nwell}$ was set to the 0.8\,V and a 1 MIP LET was considered. Even at thicknesses as large as 300\,$\mu$m, $t_{95}^{central}$ does not exceed 6\,ns. In the best-case scenario without the deep p-well, reducing the sensor thickness from 50\,$\mu$m to 40\,$\mu$m, 30\,$\mu$m and 20\,$\mu$m results in a decrease in $t_{95}^{central}$ of 15\%, 33\% and 50\%, respectively. Analogous proportionality was observed for $t_{95}$. Therefore, for future production runs, thinner sensors could be considered for the enhancement of the timing performance.

\subsection{Charge sharing}
\label{subsec:charge_sharing}

A set of TCAD simulations was dedicated to study the charge sharing among adjacent microstrips when particles with different LETs traverse the sensor. Charge sharing is relevant for improving the spatial resolution, especially with analog readout, and is enhanced by fine microstrip pitches and large sensor thicknesses. On the contrary, it is reduced at higher $V_{back}$ for a fixed sensor thickness.

\begin{figure}[H]
\centering
\includegraphics[width=9.2 cm]{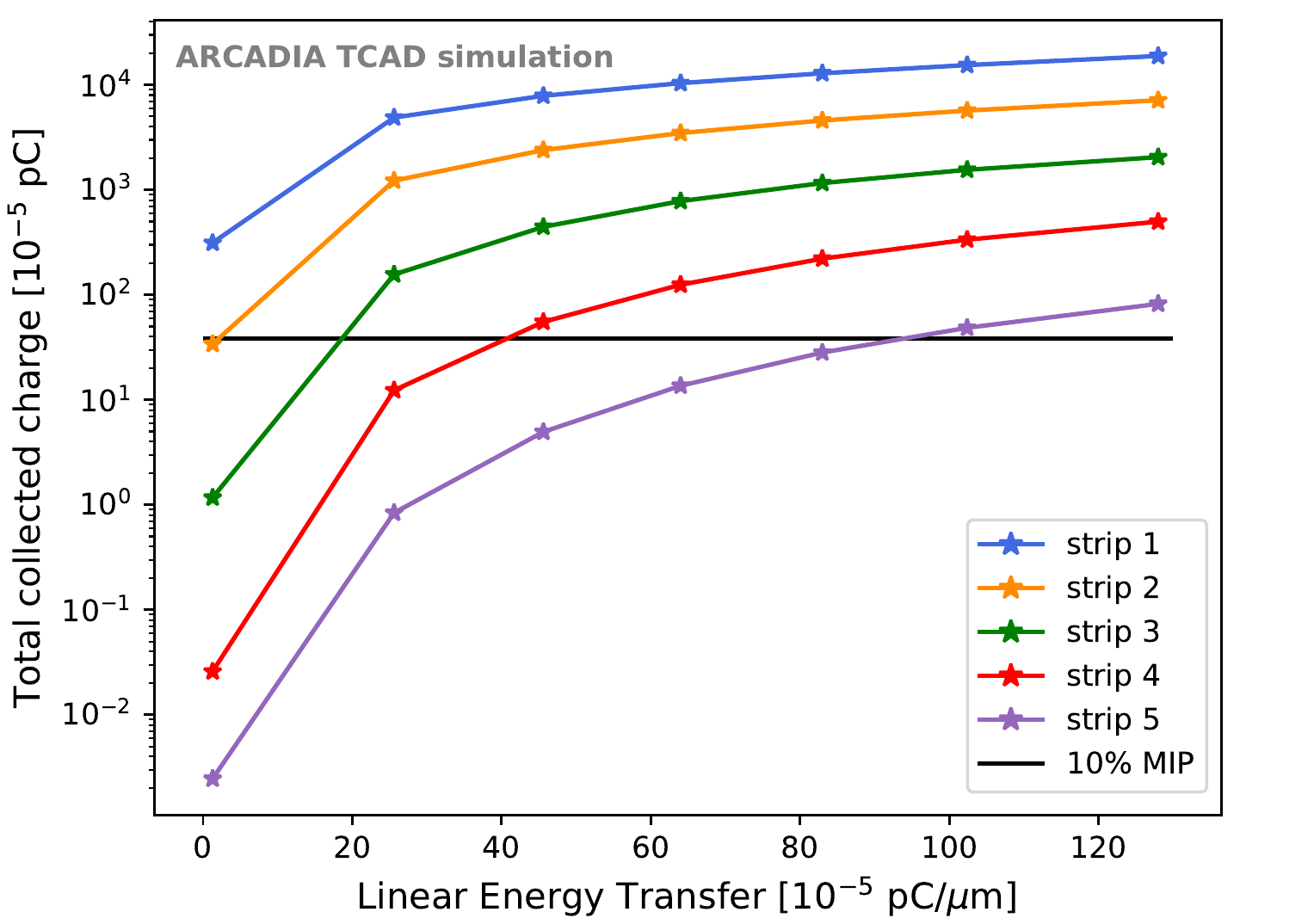}
\caption{Charge sharing among adjacent microstrips. The total charge collected by strips 1 to 5 (following the nomenclature illustrated in Figure~\ref{fig:best_worst}) is shown as a function of the LET.}
\label{fig:charge_sharing}
\end{figure}

In Figure~\ref{fig:charge_sharing}, the case of a 300\,$\mu$m thick sensor at $V_{back} = V_{pt}$ is presented for the best-case scenario. The total charge collected by each strip (identified using the nomenclature of Figure~\ref{fig:best_worst}) is plotted versus the LET. The black horizontal line indicates a possible charge threshold corresponding to 10\% of a MIP at the single strip level. A comparison with the sensors that will be produced in the first ARCADIA engineering run will allow deeper investigation on the charge sharing, a fine tuning of the simulations and studies aimed at evaluating the spatial resolution of 10\,$\mu$m pitch MAMS.

\section{Conclusions}
\label{section:conclusions}

In this work, we presented detailed TCAD simulations of CMOS-based FD-MAMS, which may find use for tracking and timing in particle and nuclear physics, space and medical applications. The results of the TCAD simulation campaign, performed to design the 10\,$\mu$m pitch FD-MAMS, demonstrate their very fast and uniform charge collection, which encourages their practicality for various applications, even under heavily ionizing particles. The effect of surface ionizing radiation damage was investigated, and the layout parameters were optimized to achieve a minimum capacitance, beneficial for electronic noise reduction. The possibility to operate the sensor in full depletion and at low-power density (i.e. before the onset of the punch through current) was verified in the simulations. A preference for small collection diodes and small (deep) p-wells emerged for obtaining lower capacitance and faster sensor response. Additionally, these simulations confirmed the possibility of monolithically integrating readout architectures in the inter-strip regions for strips of 10\,$\mu$m pitch. The first FD-MAMS samples will be produced in the upcoming ARCADIA engineering production run at the beginning of 2021 and will allow the simulation results to be compared with experimental data from electrical characterisation, laser and beam irradiation tests. The promising results of the first simulation campaign on FD-MAMS will translate into further R\&D activities to enhance the sensor performance in terms of low capacitance and high timing and spatial resolution.

\appendixtitles{yes} 
\appendix
\section{Expected effects from epitaxial layer thicknesses}
\label{appendix:expected_epi_effect}

\begin{figure}[H]
\centering
\includegraphics[width=9.2 cm]{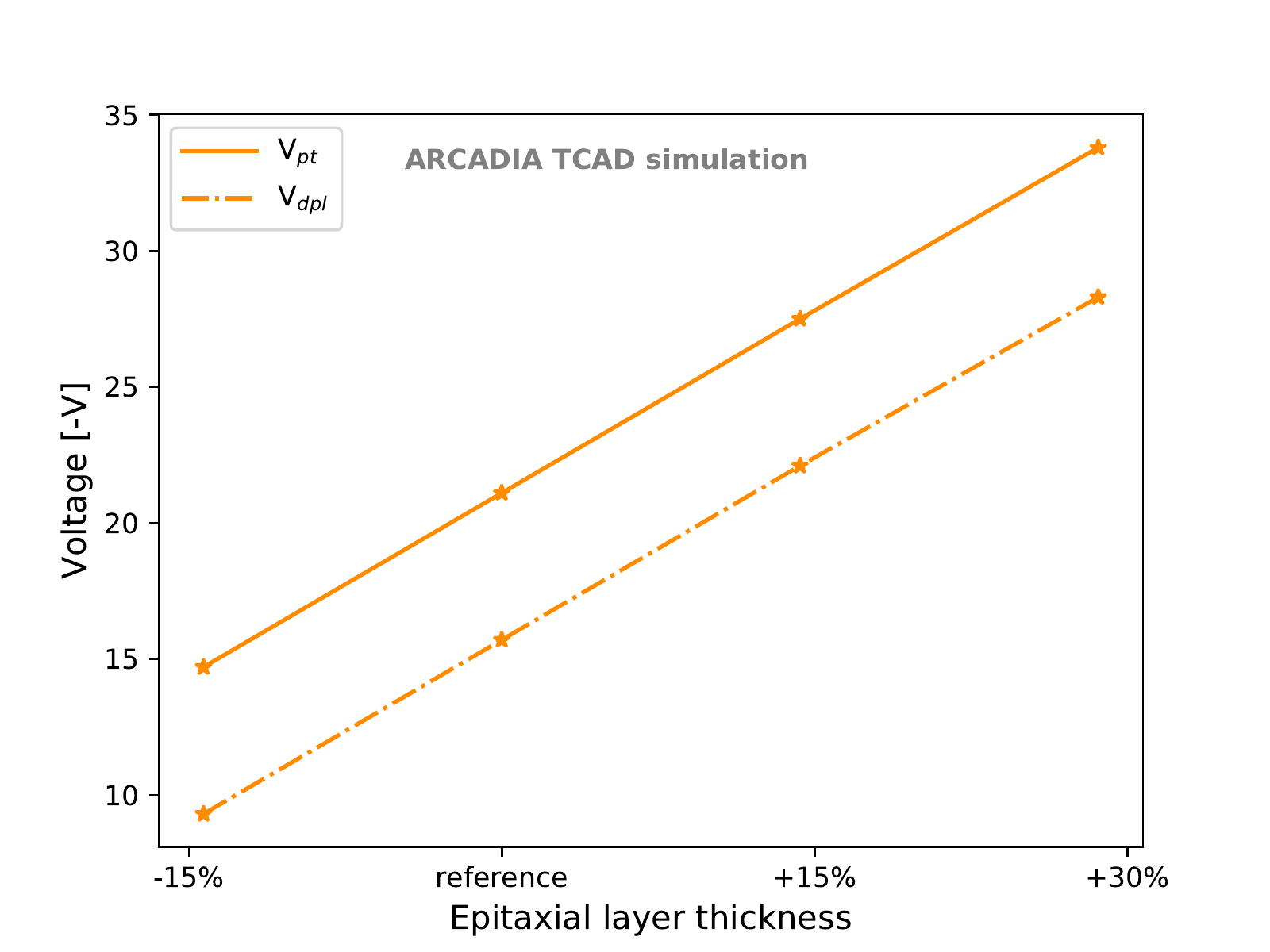}
\caption{$V_{dpl}$ and $V_{pt}$ as a function of the epitaxial layer thickness, expressed as percentage variation with respect to the reference thickness.}
\label{fig:vdpl_vpt_vs_epi}
\end{figure}

Possible variations in the epitaxial layer thickness of [-15\%; +30\%] communicated by the foundry with respect to the reference value induced us to investigate their effect on the operating parameters. While the sensor capacitance was observed not to be influenced, both $V_{dpl}$ and $V_{pt}$ showed a linear dependence on the epitaxial layer thickness. This behaviour is presented in Figure~\ref{fig:vdpl_vpt_vs_epi}.

\vspace{0.4cm}
\acknowledgments{The research activity presented in this article has been carried out in the framework of the ARCADIA experiment funded by the Istituto Nazionale di Fisica Nucleare (INFN), CSN5. The activity has also been supported by the project "Dipartimento di Eccellenza", Physics Department of the University of Torino (Dipartimento di Fisica - Università degli Studi di Torino), Italy, funded by MUR.}

\authorcontributions{Data curation, Lorenzo de Cilladi; Formal analysis, Lorenzo de Cilladi; Investigation, Lorenzo de Cilladi; Supervision, Coralie Neubüser and Lucio Pancheri; Writing – original draft, Lorenzo de Cilladi; Writing – review \& editing, Thomas Corradino, Gian-Franco Dalla Betta, Coralie Neubüser and Lucio Pancheri.}

\conflictsofinterest{The authors declare no conflict of interest.}


\reftitle{References}


\end{document}